\documentclass[apj]{emulateapj}
\usepackage{apjfonts}

\begin{document}

\title{A WIDE-FIELD {\it HUBBLE SPACE TELESCOPE} SURVEY OF THE CLUSTER Cl~0024+16
AT z=0.4. III: SPECTROSCOPIC SIGNATURES OF ENVIRONMENTAL EVOLUTION
 IN EARLY TYPE GALAXIES} \slugcomment{To Appear in ApJ}
\author{Sean M. Moran$^1$,
  Richard S. Ellis,$^1$, Tommaso Treu$^{2,3,4}$, Ian Smail$^5$,
  Alan Dressler$^6$, Alison L. Coil$^7$, Graham P. Smith$^1$ }
\affil{$^1$California Institute of Technology, Department of
  Astronomy, Mail Code 105-24, Pasadena, CA
  91125, USA email: smm@astro.caltech.edu, rse@astro.caltech.edu
  $^2$Department of Physics, University of California, Santa Barbara,
  CA 93106, email: tt@physics.ucsb.edu
  $^3$University of California at Los Angeles, Department of Physics \&
  Astronomy, Los Angeles, CA 90095
  $^4$Hubble Fellow
  $^5$Department of Physics, University of Durham, South Road, Durham
  DH1 3LE, UK.
  $^6$The Observatories of the Carnegie Institutions of Washington,
  813 Santa Barbara St., Pasadena, CA 91101
  $^7$Department of Astronomy, University of California, Berkeley, CA
  94720}
  
\begin{abstract}
We report results from a panoramic spectroscopic survey of 955
objects in the field of the rich cluster Cl~0024+1654 ($z\simeq
0.4$), complementing the {\it HST} imaging presented in the first paper
in this series. Combining with previous work, we compile a catalog
of 1394 unique redshifts in the field of this cluster, including
486 cluster members spread across an area 10~Mpc in diameter. Our
new spectroscopic sample includes over 200 high quality spectra of
cluster members. We examine the properties of a large sample of 104 cluster
early--types as a function of cluster radius and local density, using
them as sensitive tracers of the various physical processes
that may be responsible for galaxy evolution. By constructing the
Fundamental Plane of Cl~0024, we infer an evolution in the mean mass to light
ratio of early--types with respect to $z=0$ of $\Delta<Log (M/L_V)> = -0.14 \pm 0.02$. 
In the cluster center, we detect a significantly increased scatter in
the relationship compared to that seen in local clusters. Moreover, we
observe a clear radial trend in the mass to light
ratios of individual early types, with the oldest galaxies located in
the cluster core. 
Galaxies are apparently younger
at larger radius, with E+S0s in the periphery having $M/L_V$ ratios that
nearly match values seen in the field at a similar redshift. The
strong radial trend is seen even when the sample is restricted
to a narrow range in galaxy mass. Independent spectral indicators 
used in combination reveal an abrupt interaction with the cluster 
environment which occurs near the virial radius of Cl~0024, revealed by small bursts 
of star formation in a population of dim early-types, as well as by 
enhanced Balmer absorption for a set of larger E+S0s closer to the 
cluster core. We construct a simple infall model used to compare
the timescales and strengths of the observed interactions in this cluster. 
We examine the possibility that bursts of star formation are triggered when
galaxies suffer shocks as they encounter the intra-cluster medium, or by
the onset of galaxy harassment. 

\end{abstract}


\section{INTRODUCTION}

Environmental processes have clearly played a significant role in
shaping the morphological evolution of galaxies. \citet{bo78}
first noted the increased fraction of actively-star forming
galaxies in cluster cores at redshifts $z\simeq$0.4 and subsequent
studies clarified direct evolution in the relationship between
morphological fractions and local density \citep{d97}. Recent work
has extended these measures to galaxy samples in lower density
environments and at higher redshifts \citep{tt03, smith05, postman05}, 
delineating a picture
where the fraction of early-type (hereafter elliptical and S0)
galaxies to some rest-frame luminosity limit grows with time, and
at a rate that seems to depend sensitively on the local density.

What processes govern this apparent transformation of star-forming
disk and irregular galaxies into the abundant elliptical and S0
population seen in present-day clusters? Galaxy clusters provide
excellent laboratories to study these environmental effects,
particularly at intermediate redshifts  (where the relevant
processes were perhaps most active) and over the full range of
cluster radii, from the well-mixed cluster core to the outermost
regions where field galaxies are falling into the cluster for the
first time.

This is the third paper in a series concerned with exploring the
origin and evolution of the morphology-density relation via a
detailed study of hundreds of galaxies to the turn around radius
in the rich cluster Cl~0024+1654 ($z$=0.40). The essential
ingredients for this study include a large Hubble Space Telescope
WFPC-2 image mosaic providing galaxy morphologies, and Keck
spectroscopy, for membership, dynamics and diagnostics of recent
star formation activity. In the first paper in the series,  \citet[hereafter
Paper I]{tt03}, we analyzed the relation between morphology, local
density and cluster radius and pinpointed possible environmental
processes which curtail star-formation within infalling galaxies.
Local density was shown to be a more reliable measure of the
environmental trends than cluster radius, suggesting most
infalling galaxies retain their original group properties until
they reach the Virial radius where such substructure is quickly
erased. In the second paper \citep{kneib03}, a physical model for
the distribution of cluster mass  and mass to light ratio was
determined from strong and weak gravitational lensing.

In this third paper, we turn now to the spectroscopic diagnostics.
Spectroscopy of cluster members at various stages of infall can provide
a key to the dominant environmental processes. As introduced in
Paper I, different physical mechanisms will produce recognizable
spectral and dynamical signatures in the affected galaxies. While
several independent processes can operate simultaneously in
the central 0.5--1 Mpc, these can be separated by contrasting
differences over a wider dynamic range in radius and local density.

Since our first paper was submitted, we have continued to observe
Cl~0024+1654 (hereafter Cl~0024) at the Keck observatory and 
our goal here is twofold:
First, we update and finalize the redshift catalog of cluster
members. Using the DEIMOS spectrograph on {\it Keck II}, we have
now obtained spectra for nearly 1000 galaxies to a projected
distance of 5~Mpc; the tally of cluster members is nearly 500
galaxies, the largest such sample at intermediate redshift.

Secondly, in addition to measuring redshifts in abundance we have
exposed on a brighter subset of known members to a limit where our
morphological classifications are particularly reliable. At this
limit, our high-quality DEIMOS spectra can be used to examine
precise trends observed in various diagnostics of recent star
formation as well as resolved dynamics of disk and early-type
members. Our goal is to analyze these trends according to both the
timescale and physical location over which they occur in order to
develop an overall picture of the processes that affect the
cluster galaxy population.

Much of course has been learned from detailed spectroscopy of galaxies
in intermediate redshift clusters \citep[e.g.][and references
therein]{dg83, couch85,pogg99}. However, previous studies of this type
have either focused on the cluster core, where morphologies are
available or relied on spectral types or colors to trace the effects
of environment out to the cluster periphery \citep{abraham96,
kodama01}. With a larger sample of morphologies and high quality
spectra over the full range of local densities in Cl~0024, we aim to
provide a more complete picture of the environmental processes
involved and their range of application.

We focus here on cluster members classified as early-type (E+S0) in
our morphological catalog (Paper I). In the local universe,
cluster early-type galaxies are an extremely homogeneous population in
terms of their stellar populations and structural properties
\citep[e.g.][]{dressler87, dd87, bower92, bbf92}. Nevertheless, the evolution of the
morphology density relation indicates that a substantial fraction was
accreted or transformed at intermediate redshift. Several of
the proposed environmental mechanisms are thought to have transformed
spirals into S0s \citep{d97, fasano00, smith05, postman05}.

By studying these galaxies at intermediate redshifts and contrasting
their spectral properties with those of their counterparts in the
field and local universe we expect to be sensitive to
signatures of past and current environmental activity. In this
sense, we will explore early-type galaxies as ``test particles'' of
recent activity. Building on the conclusions herein, a future paper will
address the properties of spiral galaxies in Cl~0024 and discuss the
galaxy population as a whole, taking into account
morphological evolution (see, e.g., the discussion of {\it progenitor
bias} in van Dokkum \& Franx 2001).

The signal/noise of our early-type spectra was designed to be adequate
to measure reliable stellar velocity dispersions for each galaxy,
enabling us to construct the Fundamental Plane and, in particular, its
possible variation with location. This emerges as a powerful probe of
variations in the $M/L_V$ ratios and hence the luminosity-weighted
ages of the stellar populations. Precise measures of various line
diagnostics permit us to independently probe the star formation
histories over well-understood timescales. In combination, both
methods allow us to examine the relative importance of the environment
and to constrain the physical mechanisms responsible.

A key issue is the relationship between trends found in Cl~0024 at
various radii and those found in the field at approximately
the same cosmic epoch. To facilitate such a comparison we make use
of the recent comprehensive study of 163 field E+S0s undertaken by
\citet{tt05, tt05b} in the northern GOODS field.

A plan of the paper follows. In \S~2, we summarize the new
spectroscopic observations and their data reduction, and present
the final catalog of spectroscopic redshifts. In \S~3 we discuss
our measurements of the stellar velocity dispersions, the fits to
the surface photometry, the various spectral line indices, as well
as an improved estimate of the local environmental densities. In
\S~4 we present our results focusing first on the Fundamental
Plane and the implications of the scatter and various trends seen
as a function of luminosity and location, and correlations between
the Balmer absorption and metal line strengths with the velocity
dispersion. We also analyze both radial trends and those seen in
the residuals from our global cluster and field relations. In
\S~5, we develop an integrated picture which combines these
independent methods and discuss this in the light of the
conclusions we drew in Paper I. For consistency, throughout this
series of papers we use the cosmology adopted in Paper I
($H_0=65.0$ km s$^-1$, $\Omega_m=0.3$, $\Omega_\Lambda=0.7)$.

\section{OBSERVATIONS}
\subsection{Imaging}
High resolution imaging of Cl~0024 is crucial for the type of
spectroscopic study we wish to undertake, both for the purpose of selecting a
sample of E+S0 galaxies, and to allow us to analyze the surface
photometry of the selected galaxies.

Paper I presented the results of a wide-field {\it HST} imaging survey
of Cl~0024. The survey includes a sparsely-sampled mosaic of 39 WFPC2
images taken in the F814W filter ($\sim I$ band), providing good
coverage of the cluster field out to radius $> 5$ Mpc
($\sim14\arcmin$). Paper I
reported morphological classifications down to $I=22.5$.
Classifications to a limiting magnitude of $I=21.1$ were found to
be very reliable, in that several authors, working independently, 
agreed upon the morphology for most objects. This included 
differentiation between the sub-types E, E/S0, and S0. While 
ellipticals and S0s were grouped together for the purposes of Paper I, 
it is useful in this paper to detect any differences between the 
populations of Es and S0s: if spirals are actively transforming 
into S0s at $z\sim0.4$, we might expect to detect differences 
in the stellar populations of the two groups. 

Although face-on S0s are notoriously hard to distinguish from
ellipticals, especially at high redshift where S0 disks may be
too dim to detect \citep{smail97, fabricant00}, we can partially
avoid this difficulty by focusing on the brightest early type
galaxies where all but the faintest disks should be detectable.
In this paper, therefore, we will report distinctions between 
E, E/S0, and S0 galaxies for a brighter subset of this sample,
to $I=21.1$. We additionally employ a technique not discussed 
in Paper I, namely
that based on the residual signals found after subtraction
of an axisymmetric de Vaucouleurs profile ($\S$3.3 and 4.1.1).  

\subsection{Spectroscopy}

Designed to both identify cluster members and acquire high
signal to noise spectra of galaxies in Cl~0024, we began our
spectroscopic campaign in October 2001 with
LRIS on {\it Keck~I}. While some observations
were completed that year (see Paper I), poor weather forced us to
return in October 2002 and again in October 2003, this time making use
of the new DEIMOS spectrograph on {\it Keck~II} \citep{deimos}. The survey was
largely completed in 2003, though a small number of
additional galaxies were observed in December 2004. In total, we have
obtained new spectra of 955 objects, including 261 confirmed cluster
members to $I=22.5$.

Spectroscopic targets were selected from the CFHT
$I$--band mosaic of the Cl~0024 field \citep{czoske}.
In designing slit masks, selection priority was given to known
cluster members with {\it HST} morphologies, followed by galaxies in
the {\it HST} survey without a known redshift, to $I=22.5$.
Galaxies without {\it HST} images filled the remainder of each slit
mask, with priority
given to known members.  All targeted galaxies were brighter than
$I=22.5$. Masks were designed so
as to provide good coverage across the Cl~0024 field, while also
maximizing the number of known-member spiral galaxies that could
be observed with tilted slits (for extracting rotation curves from
extended line emission; to be discussed in a later paper.)

\begin{figure}[t]
\centering
\includegraphics[width=\columnwidth]{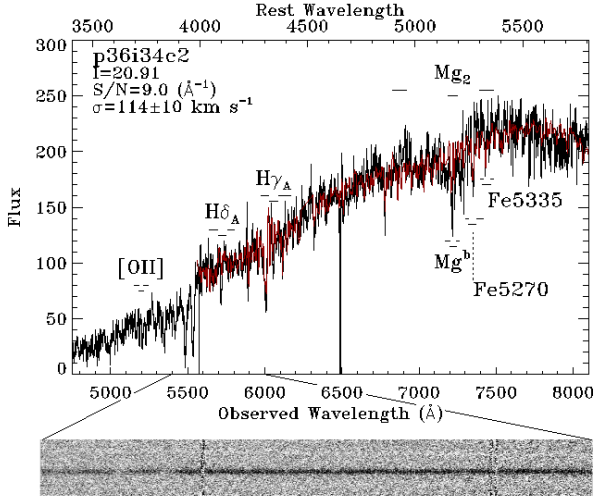}
\caption{\label{fig:example} A typical spectrum for a galaxy near the
  magnitude limit of our brighter sample. The entire 1D
  spectrum is plotted at top, with a segment of the 2D spectrum
  displayed below it. We have measured equivalent widths for the
  spectral lines indicated. The pseudo-continuum and index bandpasses
  used for each index are marked. Overplotted in red is the best-fit
  template spectrum used in measuring the stellar velocity dispersion,
  for the full spectral range where it overlaps with our galaxy
  spectrum. The resulting velocity dispersion is
  $\sigma=114\pm10$~km~s$^{-1}$.}
\end{figure}

We observed a total of 12 DEIMOS slit masks in 2002 and 2003, covering the
entire field of the HST mosaic. We employed the 900 line/mm grating,
with a central wavelength of 6200$\mbox{\AA}$. This setup provides spectral
coverage from 4500 to 8000 $\mbox{\AA}$, with a pixel scale of
approximately $0\farcs12 \times$ 0.44 $\mbox{\AA}$.
Slitlets were milled to be $1\arcsec$ wide, providing spectral resolution
of $\sigma \sim 50$km~s$^{-1}$. For most masks, the exposure time was
2.5~hrs (5 $\times$ 1800s), though four masks were only observed for
2~hrs (4 $\times$ 1800s). In December 2004, we observed a single
additional mask for 3~hrs, 10 min, (5 $\times$ 1200s plus 3 $\times$
1800s) with the 600 line/mm grating, providing
resolution of $\sim70$km~s$^{-1}$. The central wavelength was again
6200$\mbox{\AA}$, providing similar spectral coverage. In 2002,
conditions were fairly poor,
with thin clouds frequently interrupting observations. Seeing was
approximately $0\farcs7$.
In 2003, seeing was good (0\farcs5--0\farcs6), though conditions were
not photometric. And in 2004, conditions were generally good, with
seeing varying between 0\farcs7--1\farcs1 across three nights.

We analyze data for 104 E+S0 galaxies, of which 71 have particularly 
high quality spectra (generally defined in this paper to be those for
which reliable stellar velocity dispersions were measured). 
Twelve of these E+S0s (6 high quality) come from
the LRIS observations discussed in Paper I, with the remainder from
DEIMOS. Broken down by specific morphological
type, our
sample includes 34 galaxies classified as E, 50 as S0, and
20 as E/S0. Of the galaxies with high quality spectra available, 27
are E, 38 S0, and 10 E/S0.

We designed our spectroscopic campaign to yield high quality
spectra for galaxies brighter than $I=21.1$, in order to match the
magnitude limit of Paper I for precise morphological
classification. As expected, all objects where we have obtained
high quality spectra are brighter than this limit. We therefore
divide our E+S0 galaxies into a sample that is brighter than
$I=21.1$ and one that includes all observed E+S0 members down to
$I=22.5$. For the larger sample, signal/noise is generally
sufficient to measure and examine trends in spectral line
strengths. An $M_V^*$ galaxy in Cl~0024 corresponds to $I=19.5$
\citep[Paper I]{smail97}, such that our two samples represent
galaxies brighter than $M_V^*+1.6$ and $M_V^*+3.0$, respectively.

\subsubsection{Data Reduction}
Spectra were reduced using the DEEP2 DEIMOS data
pipeline\footnote{Software available at
  http://astron.berkeley.edu/$\sim$cooper/deep/spec2d} \citep{davis03}.
The
pipeline performs bias removal, flat-fielding, and cosmic-ray
rejection. It then separates slitlets and performs wavelength
calibration and sky subtraction. For wavelength calibration, the
pipeline uses an optical model of the DEIMOS mask to generate an
initial wavelength solution. Arc-lamp frames, consisting of 1s
exposures with Ne, Ar, Kr, and Xe lamps, are then used to refine
the calibration.

From the reduced two-dimensional spectra, the pipeline extracts
one-dimensional spectra using either a variance-weighted boxcar function,
or a variant of the optimal extraction method described by
\citet{horne}.
We perform our analysis on spectra extracted via the boxcar
method, though the difference using the optimal extraction is minimal. Figure~\ref{fig:example} presents a 1D reduced
spectrum of a galaxy that is near the magnitude limit of our brighter sample, along with a portion of the reduced 2D spectrum for the
same object.

For LRIS data, individual slitlets were separated and reduced in a
standard manner \citep[see, e.g.,][]{tt99, tt01b}.

For each reduced spectrum, redshifts were determined by examining
the spectra to identify key absorption and emission lines. We
obtained initial redshift guesses via automated fitting of
template stellar spectra to each object spectrum. Each spectrum
was then judged by eye to determine the correct redshift. Most
redshifts were identified by SMM, and checked by TT. Table~1 lists the
basic properties of the sample of 104 E+S0 galaxies. In later tables,
galaxies will be referred to by the object name listed
here.

\subsubsection{Cl~0024 Redshift Catalog}

\begin{figure}
\centering
\includegraphics[width=\columnwidth]{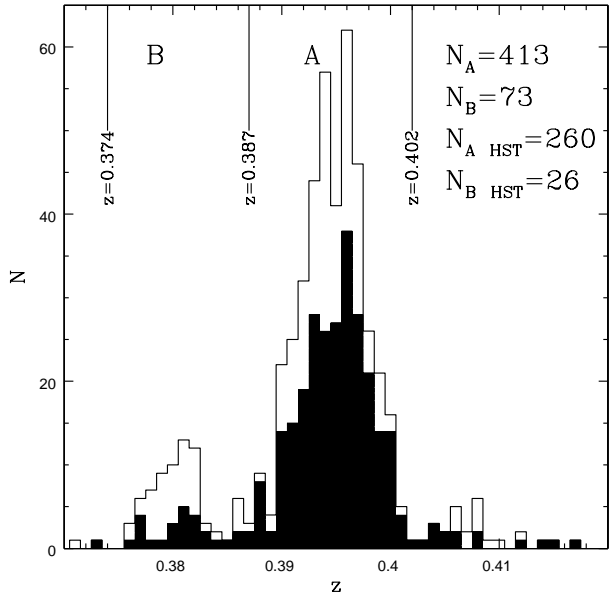}
\caption{\label{fig:redshifts} Redshift distribution of galaxies in
  the field of Cl~0024 in the vicinity of $z\sim0.4$. From Paper
  I, updated to include the final tally of redshifts in our Keck
  spectroscopic campaign. The empty histogram reflects all known
  members, while the shaded histogram represents members present in
  the HST mosaic of Paper I. The two peaks in the distribution are
  discussed by \citet{czoske}.}
\end{figure}

As in Paper I, we combine our new spectroscopy with data from
previous surveys \citep{czoske01, d99} to compile a comprehensive
redshift catalog of objects in the Cl~0024 field. In total, we
have compiled a catalog with 1394 redshifts of unique objects, of
which 486 are cluster members. Table~2 presents an excerpt from the 
total redshift catalog, which is available in its entirety online. 
Approximately sixty unique
redshifts were kindly provided by D. Koo \& A. Metevier. These are omitted
from Table~2 (reducing the tally to 1334 objects) and will be 
published separately by Metevier et al. (2005, in preparation).
A final merged catalog will be available from the two groups'
websites. As in Paper I and \citet{czoske},
we define a cluster member to be a galaxy that lies in either Peak
A or Peak B of the cluster, as illustrated in
Figure~\ref{fig:redshifts}. This encompasses a redshift range from
$z=0.374$ to $z=0.402$.

\begin{deluxetable}{cccccccc}
\centering
\tablewidth{0pt}
\tablecaption{Cl~0024+1654 Redshift Catalog}
\tablenum{2}
\tablehead{\colhead{$\alpha$} & \colhead{$\delta$} &
  \colhead{$z_{best}$} & \colhead{Quality\tablenotemark{a}} &
  \colhead{Source\tablenotemark{b}} & \colhead{$\left< z \right>$} & \colhead{$\delta z$} & \colhead{$N_z$} \\
\colhead{($^o$)} & \colhead{($^o$)} & \colhead{} & \colhead{} & \colhead{} & \colhead{} & \colhead{} & \colhead{} }

\startdata
6.845741 & 17.133789 & 0.3940 & 0 & 2 & 0.3940 & 0.0000 & 1 \\
6.837580 & 16.997200 & 0.3966 & 1 & 1 & 0.3964 & 0.0004 & 2 \\
6.827622 & 17.378466 & 0.3792 & 0 & 2 & 0.3792 & 0.0000 & 1 \\
6.821529 & 17.200779 & 0.3758 & 0 & 2 & 0.3758 & 0.0000 & 1 \\
6.812725 & 17.213043 & 0.3813 & 0 & 2 & 0.3813 & 0.0000 & 1 \\
6.801040 & 17.199699 & 0.3955 & 1 & 5 & 0.3955 & 0.0001 & 2 \\
6.765210 & 17.191099 & 0.3955 & 1 & 5 & 0.3956 & 0.0002 & 3 \\
6.759056 & 17.073933 & 0.3790 & 0 & 2 & 0.3790 & 0.0000 & 1 \\
6.744001 & 17.070641 & 0.3940 & 0 & 2 & 0.3940 & 0.0000 & 1 \\
6.737525 & 17.066620 & 0.1870 & 2 & 3 & 0.1870 & 0.0000 & 1 \\
... \\
\enddata
\tablenotetext{a}{Quality codes: $0=$ Quality unspecified by source, $1=$ Secure, $2=$ Probable, $3=$ Uncertain}
\tablenotetext{b}{Source codes: $1=$ \citet{czoske01}, $2=$ Frazier
  Owen (private communication),
  $3=$ Hale/COSMIC, $4=$ Keck/LRIS, $5=$ Keck/DEIMOS}
\tablecomments{The complete version of this table is in the electronic edition of
the Journal. The printed edition contains only a sample. Explanation
of columns: $z_{best}$, Quality, and Source refer to the most reliable
redshift for each object. If a redshift is available for an object
from more than one source, then $\left< z\right>$ gives the mean
redshift from all sources, $\delta z$ is the rms difference between
them, and $N_z$ is the number of redshifts included in the mean. }

\end{deluxetable}

While Table~2 includes galaxies from across the entire field of Cl~0024,
our sample of E+S0 galaxies is drawn from the area imaged by
WFPC2, so we are particularly concerned with the distribution of
known redshifts within the area of the {\it HST} survey.  To draw
firm conclusions about the early type population of Cl~0024, we
must draw our E+S0 galaxies from a sample that is sufficiently
complete and representative of the population as a whole.

\begin{figure}[ht]
\centering
\includegraphics[angle=270, width=\columnwidth]{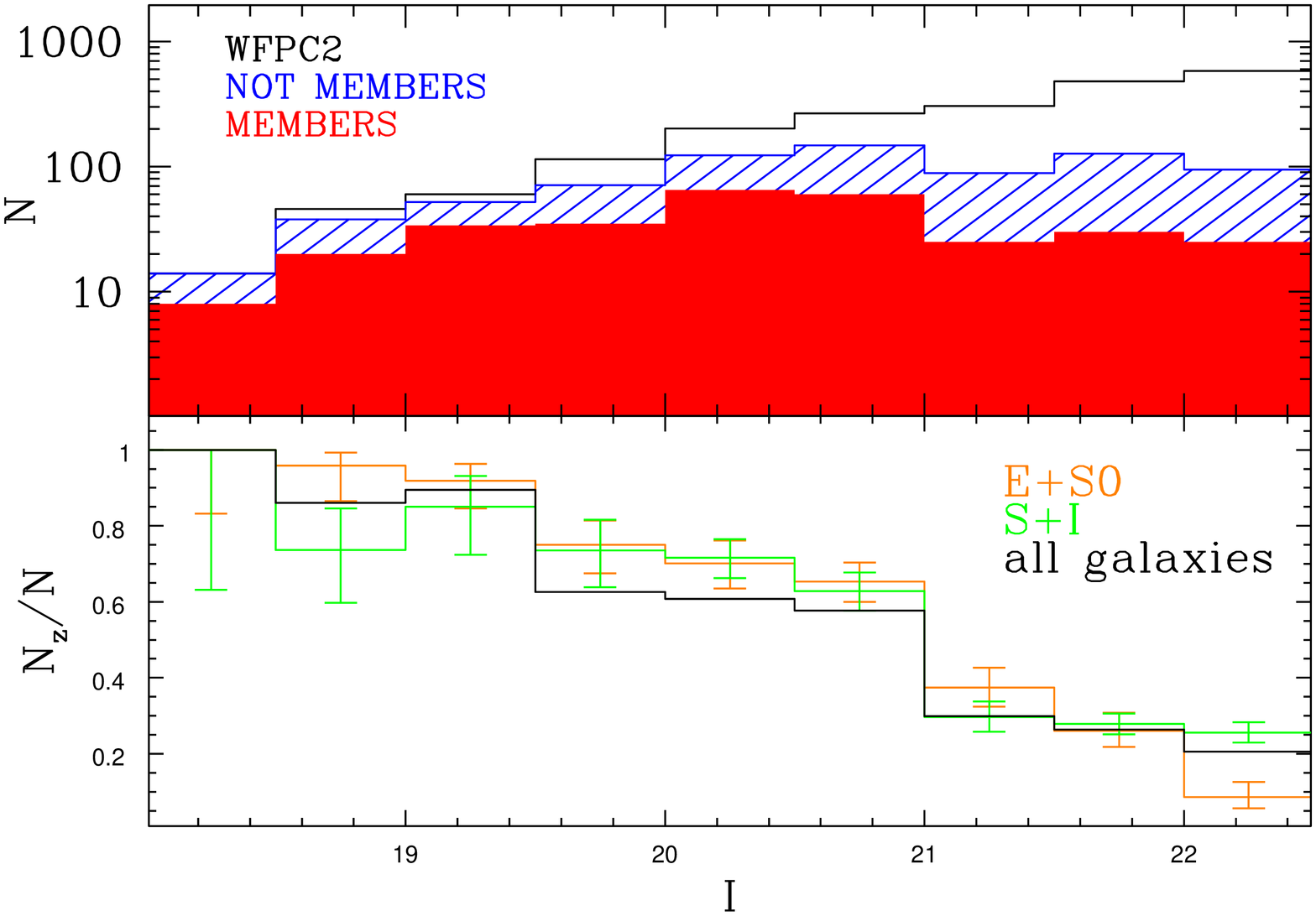}
\\
\includegraphics[angle=270, width=\columnwidth]{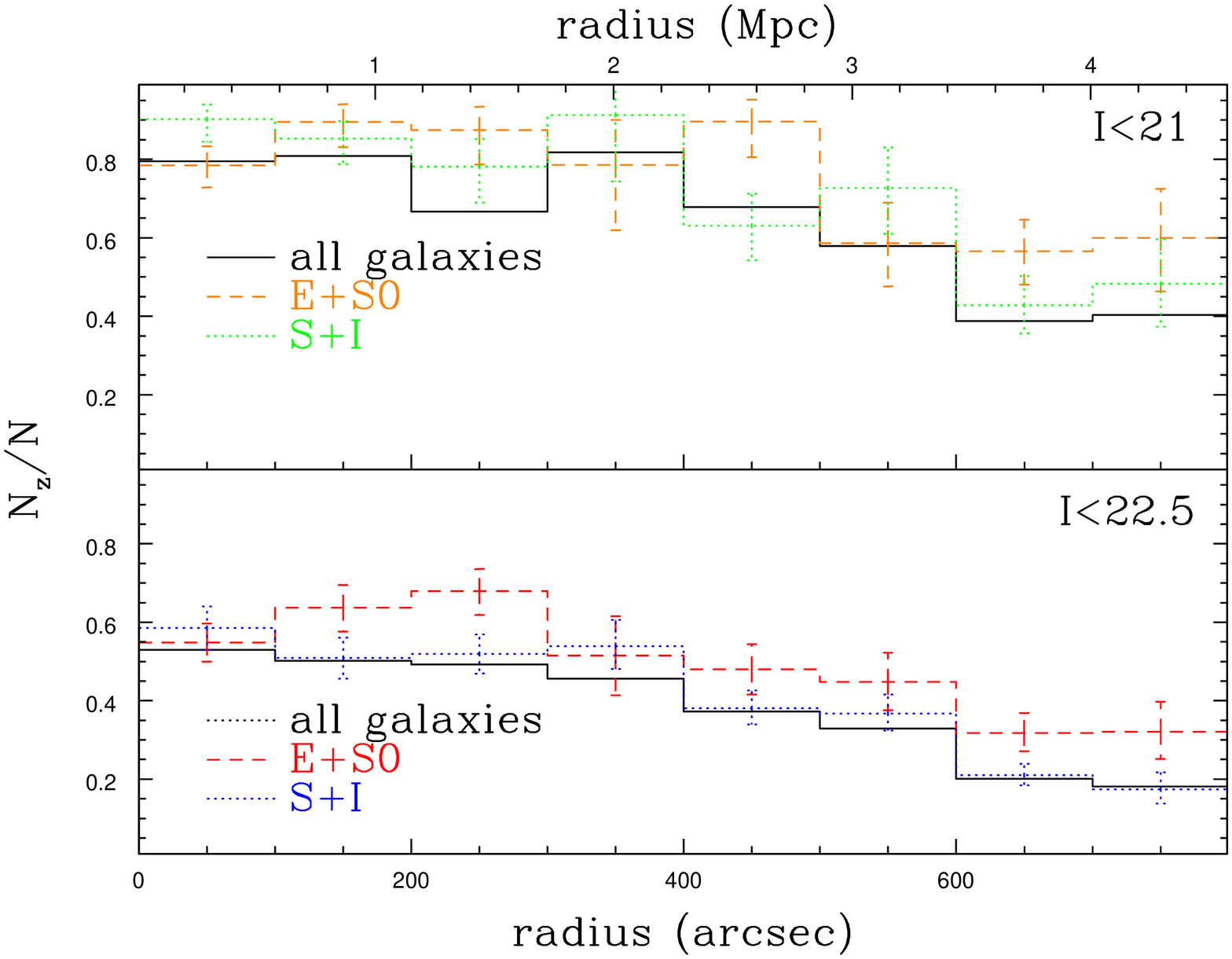}
\caption{\label{fig:completeness} Top: Number counts as a function 
of I magnitude for all objects
observed by WFPC2, with the number of cluster members and nonmembers
marked (upper panel). For all objects with WFPC2 imaging, the lower
panel displays the fraction with a measured redshift, as a function of 
I magnitude, and divided by morphology as indicated.
Bottom: Fraction of objects with measured redshift,
as in the top plot, but as a function of cluster radius, and to
two separate magnitude limits, $I<21.1$ (top) and $I<22.5$
(bottom).}
\end{figure}

We define the redshift completeness of the
catalog, as a function of magnitude or cluster radius, as
the number of objects with identified redshifts
divided by the total number of objects in the {\it HST} imaging
catalog (See Paper I). 
Completeness has naturally increased from the values given in
Paper I; ignoring objects brighter than the brightest cluster galaxy
($I=17.75$), it is now $\sim 65\%$ for $17.75<I<21.1$ and
$\sim 40\%$ for $17.75<I<22.5$. 
Importantly for this work, our coverage of E+S0 galaxies
within 3~Mpc of the cluster center is particularly
high. Within this radius, Figure~\ref{fig:completeness} shows that we
have identified redshifts for
over 80\%  of E+S0 galaxies brighter than $I=21.1$. At larger radii,
and for dimmer E+S0s ($I<22.5$), completeness drops to
$\sim50$\%. Spiral galaxies show a similar level of completeness, and
so we reaffirm the conclusion of Paper I that there is no significant
morphological bias in our redshift catalog (See Figure~\ref{fig:completeness}.)
Our sample of E+S0 galaxies should be representative of the overall
population, since the selection of galaxies for the spectroscopic
survey was largely random, and the quality of the observed spectra
depends only on random factors such as weather, introducing no bias with
respect to spatial distribution.

\section{ANALYSIS}
The powerful combination of {\it HST} imaging and high-quality DEIMOS
spectroscopy enables us to combine measures of the kinematic and
photometric structure of cluster early types with detailed spectral
information that reveals the underlying stellar population. Locally,
early type galaxies show several tight correlations between kinematic,
photometric, and spectral properties, such as the Fundamental Plane
(FP) \citep{dd87}, the Mg--$\sigma$ relation \citep{bbf93}, and the
Balmer--$\sigma$ relation \citep{kunt00, kelson01}. Examining these
relations at $z\sim 0.4$ not only gives insight into the redshift
evolution of cluster early types, but, by examining the residuals from
these relations as a function of cluster environment, we can uncover
subtle trends in the mass to light ratios ($M/L_V$), metallicities,
and star formation histories of these galaxies.

Of course, it is important first to clearly define meaningful
measures of the cluster environment of a galaxy. Following Paper
I, we measure radius as the distance from the center of the
cluster's x-ray emission. Our results would be negligibly affected
if we instead adopted the center of mass of the system 
determined by \citet{kneib03}, as the
two positions are separated by less than $5\arcsec$ ($<30$~kpc).  
In order to more reliably study
variations in galaxy properties as a function of local density, we
re-calculate the local densities presented in Paper I, now making
use of our extensive spectroscopic catalog, supplemented by the
photometric redshift catalog of \citet{smith05}.

Stellar velocity dispersions serve as a tracer of galaxy mass, and
allow us to quantify how stellar populations vary with a galaxy's
size (and, by extension, the formation history of the galaxy.) The
redshift evolution of early type galaxies can be constrained by
comparing the tight $\sigma$ correlations observed locally to what
is observed at higher redshift. We can also use velocity
dispersions to determine if trends in stellar populations are more
tightly correlated with environment or with galaxy mass.

With the addition of surface photometry from the {\it HST} images,
we can derive the parameters of the fundamental plane (FP) in
Cl~0024: effective radius $R_e$, mean surface brightness $\left< \mu_V
\right >$, and central velocity dispersion $\sigma_0$
\citep{dd87,dressler87}. The residuals from the FP reveal variations
in the mass to light ratios of galaxies; from this, we can trace how
the luminosity-weighted ages of early types vary across the cluster
environment. We can also measure the evolution of $M/L_V$ with
redshift, and examine whether the amount of this evolution depends
on galaxy mass and/or cluster environment.

We also wish to directly examine environmental trends in spectral line
strengths, as these can reveal trends in current star
formation (via [OII], [OIII],
or H$\beta$ emission), recently completed star formation (via H$\gamma$ and
H$\delta$ absorption line strengths), or metallicity (via Mg$_2$, Mg$^b$ or
the composite index $[MgFe]^\prime$).

Below, we discuss the methods we follow to accurately measure each
of the quantities necessary for our analysis: local density,
velocity dispersion, surface photometry, and spectral line indices.

\subsection{Local Density Measurements}
Local density measurements in Paper I relied on a statistical
field subtraction, following the methods of previous work
\citep[e.g.][]{wgj93,d97}. While adequate, we can improve on these
measurements by making use of our large spectroscopic catalog,
supplemented by the extensive catalog of photometric redshifts
from \citet{smith05}. These catalogs allow us to eliminate most 
foreground and background galaxies, and calculate local densities
based only on the positions of confirmed or possible cluster members. 
A standard method of measuring local
density, used in Paper I and first introduced by \citet{d80}, 
involves calculating the area enclosed
by the ten nearest neighbors of a galaxy (to $I=21.1$ in the case
of Paper I). 
In order to obtain local density measurements for our
entire sample of 104 galaxies ($I\le22.5$), we modify the method from 
Paper I to include all fainter galaxies to $I=22.5$ in the tally 
of nearest neighbors. Our method is as follows:

Each object with $I<22.5$ is given a weight between zero and one,
according to the procedure outlined below. Then, for each object, we
calculate the total area enclosed by a set of neighboring galaxies
whose cumulative weight equals about 10. (Fractional weights are
common, so the total weight rarely equals exactly 10.)
The local density is then calculated by dividing the total weight
by the area in Mpc$^2$. Compared to the method used in Paper I, this
should give a value of $\Sigma_{10}$ that is a better reflection of the true
density of galaxies in the cluster, minimizing errors due to chance
superpositions of background galaxies or groups.

As we wish to include only cluster members in the calculations of
local density, we assign weights to galaxies based on how confident we
are that it is a cluster member. If a spectroscopic
redshift is available from the combined catalog (\S~2.2.2), then
the object's weight equals 1 if it is a cluster member
($0.374 < z < 0.402$), and zero otherwise.

For objects with no spectroscopic redshift, we check for a
photometric redshift in the catalog of \citet{smith05}. While the
cluster is clearly evident in the redshift distribution of
galaxies in the \citet{smith05} catalog, the limited accuracy of
photometric redshifts means that the cluster galaxies are 
smeared across the redshift range $z_{phot}=0.39\pm0.09$
We give all galaxies within this range a weight of
1, even though some non-cluster members will be mistakenly included. 
From an analysis of galaxies with
both spectroscopic and photometric redshifts available, we find
that approximately $85\%$ of cluster members have photometric
redshifts in the adopted range, and about one third of all
galaxies with photometric redshifts in this range will actually
lie outside of the cluster.
 This will lead us to slightly
overestimate the total number of cluster members, and, by
extension, local densities. But the effect is minimal given the 
large fraction of spectroscopic redshifts. We also explored a more 
elaborate weighting scheme based on the joint probability that a
given galaxy will be a member according to its photometric redshift and
projected radius. The rms variation in $\Sigma_{10}$ between the
this and our simple scheme is about $12\%$ -  less than other sources 
of uncertainty and therefore not worthy of adopting.

The photometric redshift catalog is J-limited at $J\le 21.5$, so there are 
some objects with $I<22.5$ that have neither a
spectroscopic nor photometric redshift available. For these, we assign
weights according to the radial dependence of the probability that a random 
galaxy will be a cluster member. For example, based on the combined 
spectroscopic redshift catalog, the probability is $81\%$ that an unknown 
galaxy within 0.75 Mpc of the cluster core is a cluster member; such a
galaxy is assigned a weight of 0.81. We calculate and assign similar
probabilities for galaxies in several different radial bins out to
5~Mpc. Since we only consider galaxies within 
the magnitude range $17.75\le I < 22.5$, the probabilities we have 
adopted do not vary strongly with magnitude, and so we do not further 
subdivide our probability estimates into different bins for different 
ranges of galaxy magnitude.

In order to properly compare densities calculated with our method to
those used in Paper I,  we also calculate $\Sigma_{10}$ using objects 
brighter than $I=21.1$. A tight correlation is seen between the
two estimates which agree to within 20\%;  we also  reproduce the 
overdensities at $\sim1$~Mpc and $\sim3$~Mpc given in Paper
I. Similarly, we can compare the density of background objects
predicted by our method to the field number counts of, e.g.,
\citet{abraham96a} and \citet{postman98}. We calculate background
count densities of $\log(N)/deg^2=4.45\pm0.05$ (to $I=22.5$) 
and $3.90\pm0.07$ (to $I=21.1$). Our predicted counts 
agree with both \citet{abraham96a} and \citet{postman98}, 
within the uncertainties, for both magnitude 
limits. As an additional check on the uncertainty in $\Sigma_{10}$ measured 
to our deeper limit ($I=22.5$), we calculated a density ($\Sigma_5$) 
for a total weight of  5. The rms variation between $\Sigma_5$ and
$\Sigma_{10}$ is about $25\%$. Conservatively, we adopt this as 
the uncertainty in $\Sigma_{10}$. 

\subsection{Stellar Velocity Dispersions}
We are able to measure velocity dispersions only for our brighter
sample of early-type members ($I<21.1$), as our spectra of fainter
galaxies do not have sufficiently high signal to noise.
In order to determine velocity dispersions, we fit to a grid of
stellar templates degraded to the instrumental resolution and
smoothed to various velocity dispersions \citep{pixfit}. A high 
quality spectrum for an object near our magnitude limit is 
plotted in Figure~\ref{fig:example}, with
the best-fitting template spectrum overplotted.

To determine the signal/noise limit at which our velocity
dispersion measures become unreliable, we performed a series of
Montecarlo simulations. We construct fake galaxy spectra from
stellar templates smoothed to the resolution and pixel scale of
DEIMOS (for the 900 line/mm grating), truncated to an identical
length of $\sim2600\mbox{\AA}$, convolved with a Gaussian of
various widths to simulate different velocity dispersions, 
and degraded to a variety of signal to noise ratios. We then 
attempt to recover the velocity dispersion of the fake galaxy 
by running the same code as above.

We find that the approximate mean $S/N$ where systematic errors in 
$\sigma$ reach $\sim10\%$ corresponds to  $S/N= 7-8$
($\mbox{\AA}^{-1}$, observer's frame). Below this
level, velocity dispersion measures rapidly become
inaccurate. In our high quality sample, we include galaxies with $S/N$
near this limit, though most have $S/N >10$.  However, we place a 
somewhat stricter limit ($S/N > 8$) on the spectra observed with LRIS 
and the DEIMOS 600 lines/mm grating which have slightly worse spectral 
resolution, Only 10 high quality spectra have $S/N < 10$, so our
results are fairly insensitive to these choices. 
Table~3 lists all galaxies with
high quality spectra, along with their velocity dispersions, formal
errors, and the mean signal/noise of the spectrum.

The typical uncertainty in our velocity dispersion measurements is
$\pm10\%$. These errors are dominated by differences in $\sigma$
that depend on the template spectrum used, though systematic
errors rise to become equally important as we approach the signal
to noise limit. Of the early-types where
we measured velocity dispersions, there were three galaxies in
common with an earlier study of Cl~0024 by \citet{vdf}; the
velocity dispersions quoted in \citet{vdf} match ours in all three
cases, with $<\delta\sigma/\sigma>= -0.02\pm0.13$.
\citet{tt05b} derived stellar velocity dispersions
from DEIMOS spectra using a similar method to our own. They
pursued several tests to determine the accuracy of their
dispersions, and found an rms uncertainty of $\sim12\%$, in
agreement with our own uncertainty estimates. For more discussion of
such accuracy tests, see \citet{tt05b}.

For each galaxy, we apply a correction to match the central
velocity dispersion measured through a $3.4''$ aperture at the
distance of Coma, following the prescription of  \citet{jfk95a}.
This choice of aperture size
for the correction is somewhat arbitrary, but is a common choice
for studies of early-type galaxies at low to intermediate redshift
\citep[e.g.][]{kelson00c, wuyts04} because it facilitates comparison to local
measurements of the Fundamental
Plane \citep[e.g.][]{jfk96}.  The magnitude of this correction
depends on the physical scale over which the 1D spectrum was
extracted, which varies from object to object. The average
correction applied is $6.6\% \pm 0.4\%$. Corrected velocity
dispersions are denoted by $\sigma_0$, and are listed in Table~3.

\subsection{Surface Photometry}

{\it GALFIT} \citep{peng} was used to derive effective radii and
surface brightnesses for all galaxies with measured velocity dispersions.
For each galaxy, we extract postage stamps approximately 200
pixels on a side ($\sim 10\arcsec$). We then use {\it GALFIT} to fit against a
model de Vaucouleurs ($r^{1/4}$) profile. Following
other authors, e.g. \citet{jfk95}, we fit all E+S0s to the same
profile shape, even though some may be better described by a lower
S\'{e}rsic index, or a de Vaucouleurs plus exponential function. 
(See \citet{peng} for a definition of the S\'{e}rsic function, a more 
general form of the de Vaucouleurs function.) 

{\it GALFIT} minimizes the $\chi^2$
residuals between the galaxy image and a 2D galaxy model that it
constructs. The free parameters in this model
profile include: $R_e$, total magnitude, axis ratio, galaxy position
angle, the position of the galaxy center, and the sky level. Sky
levels were set according to the header of each WFPC2 image, but were
allowed to vary within a small range to ensure that {\it GALFIT}
converges to the correct fit.

Model galaxies are convolved with a PSF
before fitting; we use a star observed on the same WFPC2 chip, with
approximately the same exposure time. We tested with a number of
different stars, and found that the specific choice of PSF star
did not significantly affect the derived photometric parameters.

When fitting a galaxy to a de Vaucouleurs profile, the best-fit
parameters are particularly sensitive to
extra flux far from the galaxy's center \citep{peng}, since the function
declines relatively slowly beyond $R_e$. Neighboring or overlapping
galaxies thus contribute light that must be either masked or removed
by fitting multiple galaxy profiles. To best remove this extra light, we
simultaneously fit a S\'{e}rsic profile to each bright neighbor within
the postage stamp image. The S\'{e}rsic function is the best choice
for such a fit because its form is general enough to successfully model a wide range of
galaxy types; according to \citet{peng} this ensures that the galaxy's
flux is subtracted uniformly.

\begin{figure}[t]
\centering
\includegraphics[width=\columnwidth]{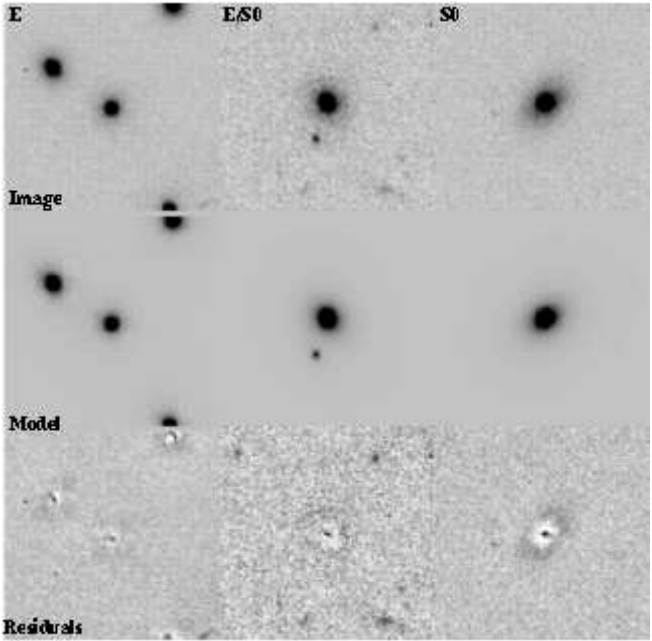} \caption{\label{fig:galfits}
  Example surface photometry fits, for galaxies classified as
  E, E/S0, and S0 by Paper I. Top row is the galaxy image. Middle row
  shows the GALFIT model image. Bottom row shows the residuals of the
  image fit to model. }
\end{figure}

The magnitude of the best-fit model galaxy returned by {\it GALFIT}
is measured in the observed
F814W filter. At $z\sim0.4$, this is a close match to rest-frame V,
but we must still apply a small {\it k--color correction} to
derive the mean surface brightness in rest frame V. $\left< \mu_V
\right>$ is defined as:
\\
\begin{eqnarray}
\left< \mu_V \right> &= I_{814W} + 2.5\log\left(2\pi R_e^2\right) +
  \Delta m_{VI} - A_I -10\log\left( 1+z \right) \nonumber \\
  &= I_{814W}+5\log\left(R_e\right)+1.29\pm0.04 \nonumber
\end{eqnarray}
\\
where $A_I=1.95 E(B-V)$ \citep{schlegel98} corrects for galactic extinction, the redshift
term accounts for cosmological dimming, and $\Delta m_{VI}$ is the {\it
  k--color correction}. In the second line of the above equation, we insert our adopted
values: $\Delta m_{VI} = 0.85\pm0.03$, adopted from calculations by \citet{tt01},
$A_I=0.11\pm0.01$ from \citet{schlegel98}, and $z\sim0.395$.

Since we only measure surface photometry for our brighter sample of
galaxies ($I<21.1$), formal statistical errors in the measured parameters are very small: less than
$0\farcs05$ in $R_e$, and 0.05 in magnitude. We estimate that
systematic errors are double these values, and adopt $0\farcs1$ and
$0.1$ mag as typical errors in $R_e$ and magnitude,
respectively. There may be additional uncertainty in $R_e$ and $\mu_V$ 
related to the choice of a de Vaucouleurs profile over other
structural forms, but previous work \citep{fritz05,kelson00a,saglia93} has shown
that the combination of $R_e$ and $\mu_V$ that enters into the
Fundamental Plane (see \S~4) is largely insensitive to the galaxy profile adopted.
Figure~\ref{fig:galfits} shows three example fits, for galaxies
classified as E, E/S0, and S0. For each galaxy, we display the
original galaxy image, the best-fit model image, and the residuals. As
might be expected, the residuals are smaller for the fit to the
elliptical galaxy; the residuals for the S0 galaxy clearly show a disk
component that is not well fit by a de Vaucouleurs profile.

We observed two clear edge-on S0s which had to be removed from our
sample of high-quality spectra, due to the uncertainty in trying
to fit a de Vaucouleurs profile to such an edge-on disk. We also
removed two galaxies with bad fits, defined as where the
SExtractor magnitude from the {\it HST} image and {\it GALFIT} model
magnitude differ by more
than 0.75--the two magnitudes for most galaxies in our sample match to much
better than 0.75 magnitudes. None of these
four galaxies are included in the previously defined sample of 71
high-quality early-types, though they are included in the larger
sample of 104 galaxies. The photometric parameters for the
high-quality galaxies are listed in Table~3.

As a check on our measurements, we compared our $R_e$ and $\left< \mu_V
\right>$ for three galaxies that also were studied in the
work by \citet{vdf}. (This is not the same set of three we used to
compare velocity dispersions in \S~3.2: for one galaxy where each
group has measured surface photometry, we have not measured $\sigma$. And
for one galaxy where we have both measured $\sigma$, we do not have an
{\it HST} image for surface photometry.)
In each case, our surface photometry
matches theirs, within the adopted errors on our measurements.
We exclude one additional galaxy
that we have in common with \citet{vdf}, as it exhibits a
disturbed morphology (the triple nucleus galaxy discussed in their
paper.)

\subsection{Line Strength Measures}
We measure the strengths of several diagnostic spectral lines for the
entire sample of early-type galaxies (to $I<22.5$). In order to best
probe the stellar population of each galaxy, we select a set of
indices that are sensitive to a range of star formation histories and
metallicities. Emission lines, such as [OII], [OIII], and sometimes
H$\beta$ indicate ongoing star formation (or possibly nuclear
activity). Balmer absorption lines, such as H$\gamma$ and H$\delta$
are sensitive to recently completed star formation; these lines are
strongest in A-stars, which contribute prominently to a galaxy's
integrated starlight within the first Gyr after a burst of star
formation. We also measure several metallicity indicators, such as
Mg$_2$, Mg$^b$, Fe5270, Fe5335, and the composite index
$[MgFe]^\prime$, which is defined as:
\\
\begin{displaymath}
[MgFe]^\prime \equiv \sqrt{Mg^b\left( 0.72*Fe5270+0.28*Fe5335 \right)}
\end{displaymath}
\\
In the local universe, $[MgFe]^\prime$ seems to be insensitive to
variations in $\alpha$--element abundance\citep{thomas03}, making it
a valuable tracer of total metallicity.

Where possible, we adopt the Lick index definitions to measure the
strength of each spectral line \citep{worthey94}.  In the Lick system,
the equivalent width of a line is measured by defining a
wavelength range to either side of a main index bandpass. The mean
level of each sideband is determined, and a straight line is fit,
defining the ``pseudo-continuum'' across the index bandpass. The
equivalent width of the line within the index bandpass is then
measured with respect to the pseudo-continuum level. The Lick
system does not include an index for [OII], so we adopt the one
defined by \citet{fisher98}. In Figure~\ref{fig:example} we plot an example
spectrum with the wavelength ranges of several indices and their
sidebands marked. For clarity, the results in \S~4 will concentrate primarily on
three representative sets of measurements: [OII],
(H$\gamma_A$+H$\delta_A$), and $[MgFe]^\prime$.

Table~3 lists the strengths of several key spectral lines, for all 104 galaxies in
our sample.  These raw indices are suitable for examining environmental trends
within our own data set, but in order to make a proper comparison to
other published data or theoretical models, we must carefully
correct for any systematic differences between each set of
measurements. In particular, index strengths are known to vary with the spectral
resolution of the data. While we attempt to
compare our data only to measurements made at high spectral
resolution, to take full advantage of the high resolution available
with DEIMOS, in some cases we are forced
to degrade the resolution of our spectra to match that of the
comparison data or model.

In \S~4 and \S~5, we will compare some of our
results to the stellar population models of
\citet{bc03}, which include full synthetic spectra at a resolution of
$3 \mbox{\AA}$; this is the closest match available to the intrinsic
resolution of our DEIMOS spectra ($\sim 1\mbox{\AA}$). We will also
examine the Balmer--$\sigma$ relation in comparison to data measured
at $10\mbox{\AA}$ resolution \citep{kelson01}, the approximate resolution of the original
Lick system \citep{worthey94}. Therefore, we convolved our DEIMOS
spectra with Gaussians to produce degraded spectra at both
$3\mbox{\AA}$ and $10\mbox{\AA}$ resolutions. We then re-measured the
relevant spectral line indices. While not included in
Table~3, spectral index measurements from our degraded DEIMOS spectra
are available from the authors by request.

To compare our Balmer--$\sigma$ relation to previous work by
\citet{kelson01} on the redshift evolution of this correlation,
we apply an aperture correction to our H$\delta_A$ and H$\gamma_A$ line
strengths; we adopt their estimate that the quantity
$(H\delta_A+H\gamma_A)$ varies with aperture as:
\\
\begin{displaymath}
   \Delta \left( H\delta_A+H\gamma_A \right) = 1.78\pm0.16 \Delta \log \left(D_{ap}\right)
\end{displaymath}
\\
and correct to an aperture of $1\farcs23$ at $z=0.33$. This is a very
small correction to our data, which was measured through a similar
aperture size at slightly higher redshift.

For any fixed-width absorption line index, Doppler broadening of
lines will cause measured equivalent widths to be underestimated:
as velocity dispersion increases, more of the line's flux falls
outside of the index bandpass. We can correct for this effect by
modeling how each index varies with $\sigma$, with the help of
the \citet{bc03} population synthesis models. We select several of
their theoretical spectra at $3\mbox{\AA}$ resolution and zero
velocity dispersion, and broaden them to a series of different
velocity dispersions. At each $\sigma$, we measure the line
indices we wish to correct: $[MgFe]^\prime$, $H\delta_A$, and
$H\gamma_A$. We then fit a quadratic function in $\sigma$ to the
resulting set of measurements. We find that the $\sigma$
correction depends on the initial resolution of the spectrum, so
we also degrade the \citet{bc03} spectra to $10\mbox{\AA}$
resolution, and repeat the procedure to determine the proper
correction for our low-resolution measures of $H\delta_A$ and
$H\gamma_A$.

Since we are not able to measure velocity dispersions for
our full sample of 104 galaxies, corrected indices only appear in
plots that are limited to the 71 bright galaxies where we have accurate
measurements of $\sigma$. All plots
that demonstrate trends in the full sample of 104 E+S0s are shown with
uncorrected indices only. However, not applying a $\sigma$ correction
in these cases produces only a small error: less than
$\pm0.8\mbox{\AA}$ in $(H\delta_A+H\gamma_A)$. Emission lines like
[OII] do not vary regularly with $\sigma$, as the source of emission is
generally not spread evenly across a galaxy.

In order to compare observed spectral line strengths to the 
predictions of Bruzual \& Charlot models with various 
star formation histories, we need to calibrate
our observed [OII] equivalent widths to specific star formation
rates ($M_\odot/M_{gal}$~yr$^{-1}$). 
We make such a calibration using the deep H$\alpha$ imaging 
of Cl~0024 by \citet[][and private communication]{kodama04}. 
First, equivalent widths from the two catalogs are cross-correlated: 
our [OII] widths closely track those in H$\alpha$, approximately reproducing
the locally-observed trend \citep{kennicutt92}. For each H$\alpha$ detection,
Kodama also provide an estimated star formation rate, which we
convert to a specific star formation rate by dividing by an
estimated mass for each member. For galaxies with velocity dispersions
available, we calculate a galaxy's dynamical mass according to
$M=5\sigma^2 R_E/G$. To estimate a stellar mass for
a galaxy with no available velocity dispersion, we first determine the
typical $M/L_V$ for a local galaxy of this luminosity
from \citet{gerhard01}, and correct $M/L_B$ to $M/L_V$ by subtracting
a factor equal to $\log(M/L_V)-\log(M/L_B)=-0.06$, estimated from the
typical colors of nearby early type galaxies. 
We then correct for redshift evolution in $M/L_V$ (based on our 
Fundamental Plane results  below), and multiply by the observed 
luminosity of the galaxy:
\\
\begin{displaymath}
  \log(M)\simeq \log(M/L_V) + <\Delta \log(M/L_V)> + \log(L_V)
\end{displaymath}
\\ 
For galaxies with velocity dispersions, masses estimated in this way
are consistent with the calculated dynamical masses 
($\Delta \log(M/M_\odot)=\pm0.3$).  
We then fit a straight line to measured [OII] versus
specific star formation rate ($M_\odot/M_{gal}$~yr$^{-1}$), to
yield a conversion relationship between the two.

In order to visualize overall trends in the spectral properties of
cluster early types, we also produce a series of co-added spectra for
each radial zone. Each spectrum is normalized, shifted to the rest
frame, and then co-added. Bad pixels and sky lines are given zero
weight in the addition. This method provides a snapshot of what the
average spectrum of each ensemble of galaxies looks like. While
weighting by luminosity would better represent the integrated stellar
population of each ensemble, in practice, the co-added spectrum is
dominated by the brightest galaxy in each group. We must
be careful, however, in interpreting differences between the coadded normalized
spectra for each radial zone: each coadded spectrum will reflect an
ensemble of galaxies with a different average size and
magnitude. Therefore, it is difficult to separate radial trends in the
spectra from trends with magnitude or size.
These co-added spectra will be discussed below in conjunction with
the environmental trends in the spectral indices of individual galaxies.

\begin{figure*}[t]
\centering
 \includegraphics[width=1.5\columnwidth]{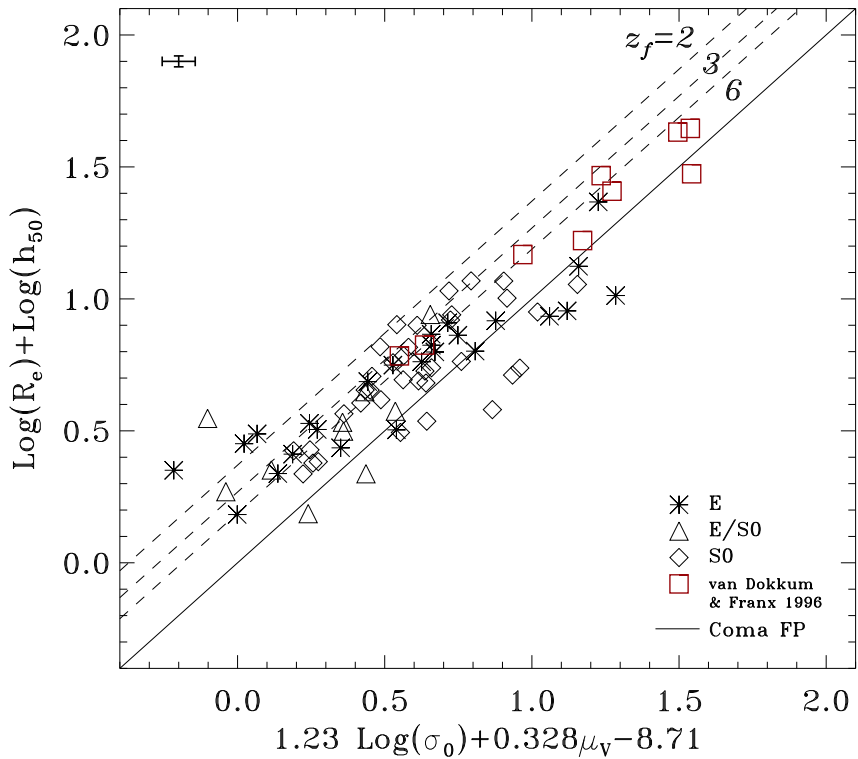}
\caption{\label{fig:fp} FP of
  Cl~0024, compared to Coma cluster (solid line). Symbols
  represent different morphologies, as indicated. 
  Dotted lines correspond to the expected
  shift in FP zero point from Coma to $z\sim 0.4$, for SSP models with
  $z_f=2.0,3.0,6.0$.}
\end{figure*}

\section{RESULTS}
\subsection{Cluster: Empirical Scaling Laws}
Before we examine environmental trends in galaxy properties, we
present the overall Fundamental Plane,
[MgFe]$^\prime$--$\sigma$ relation and the Balmer--$\sigma$ relation for the
cluster sample with high quality spectra,
and discuss how each has evolved between $z\sim0.4$ and the present epoch.

\subsubsection{The Fundamental Plane}

Previous studies have traced
a mild shift in the intercept of the cluster FP with redshift
\citep{fritz05, wuyts04,kelson00c}.
This seems to be consistent with passive
luminosity evolution of stellar populations
with a high redshift of formation \citep{wuyts04},
though biases due to morphological evolution are difficult to
quantify. However, most earlier studies have
concentrated on measuring the evolution of the FP from data taken in
intermediate or high redshift cluster cores. With our
broader spatial coverage, we can uncover any significant difference in
the mean $M/L_V$ of early types as a function of
radius. Our sample also extends to fainter magnitudes than previous
studies at $z\sim 0.4$, allowing us to probe $M/L_V$ for
smaller early-types that perhaps formed later than the most massive
cluster ellipticals.

Figure~\ref{fig:fp} presents the FP of Cl~0024 compared to that of the Coma
cluster, adopting the parameters determined locally by \citet{lucey91}:
$\alpha=1.23$, $\beta=0.328$, and $\gamma=-8.71$, where the
fundamental plane is defined as
\\
\begin{displaymath}
Log(R_e)=\alpha Log(\sigma_o)+\beta \left< \mu_V \right>+\gamma
\end{displaymath}
\\
If we assume for the moment that $\alpha$ and $\beta$ remain fixed
with redshift--i.e., that E+S0s evolve passively-- then we can
relate the offset in the intercept $\gamma$ between CL~0024 and
Coma to the change in the mean mass to light ratio of cluster
early types since $z=0.4$:
\\
\begin{displaymath}
  \left< \Delta \log \left( M/L_V \right) \right> = \left< \Delta\gamma \right>/\left( 2.5\beta \right)
\end{displaymath}
\\
We find that the average offset from the Coma FP implies a change
in the $M/L_V$ ratio between $z=0.4$ and $z=0.02$ of $\left<
\Delta \log \left( M/L_V \right) \right> =-0.14 \pm 0.02$,
excluding galaxies with velocity dispersions below $100$ km
s$^{-1}$. (Including those galaxies, the zero point shift rises to
$\left< \Delta \log \left( M/L_V \right) \right> =-0.18 \pm 0.03$)
This is a smaller evolution than found in the field at $z\sim
0.3-0.5$ by \citet{tt05b}, $\left< \Delta \log \left( M/L_V
\right) \right> =-0.23 \pm 0.05$, though it is in agreement with the
offset determined by \citet{kelson00c} for a cluster at $z=0.33$
($\left< \Delta \log \left( M/L_V
\right) \right>=-0.17\pm0.03$). Overplotted in Figure~\ref{fig:fp} are
dotted lines representing the expected evolution of the FP zero
point from the Coma cluster back to $z=0.4$. These are based on
the passive evolution of Single Stellar Population (SSP) galaxy
models, with a redshift of formation, $z_f$, of 2, 3, or 6
\citep{bc03}. The observed FP in Cl~0024 is consistent with $z_f >
3$ for most cluster early types.

\begin{deluxetable*}{rccccc}
\centering
\tablewidth{0pt}
\tablecaption{$\left< \Delta \log \left( M/L_V \right) \right>$ for
  several subsets of our data.}
\tabletypesize{\footnotesize}
\tablenum{4}

\tablehead{\colhead{} & \colhead{E} & \colhead{E/S0} &
  \colhead{S0} & \colhead{All} & \colhead{All+vDF}}

\startdata
\multicolumn{3}{l}{\bf All $\mathbf{\sigma}$\bf:}\\
N:   &      24   &   10    &    35 &    69  &  77 \\
$\left< \Delta \log \left( M/L_V \right) \right>$:& $-0.18\pm0.05$  &  $-0.23\pm0.08$ &   $-0.14\pm0.03$  & $-0.18\pm0.03$ & $-0.18\pm0.02$ \\
$\pm 1 \sigma$: &   0.23   &   0.26  &   0.20 &   0.20  & 0.19 \\
\tableline\\
\multicolumn{3}{l}{$\mathbf{\sigma > 100 km s^{-1}}$\bf:}\\
N:    &    21  &      8  &     33   &   62  &  70 \\
$\left< \Delta \log \left( M/L_V \right) \right>$:& $-0.14\pm0.03$ &
$-0.14\pm0.06$ & $-0.14\pm0.03$ & $-0.14\pm0.02$ & $-0.14\pm0.02$ \\
$\pm 1 \sigma$:  &   0.15   &  0.17  &   0.18  &  0.16  & 0.16 \\
\tableline\\
\multicolumn{3}{l}{$\mathbf{R<1}$\bf Mpc, $\mathbf{\sigma > 100 km s^{-1}}$\bf:}\\
N:    &    10  &      6  &     15   &   31  &  39 \\
$\left< \Delta \log \left( M/L_V \right) \right>$:& $-0.09\pm0.06$ &
$-0.10\pm0.06$ & $-0.06\pm0.06$ & $-0.07\pm0.03$ & $-0.09\pm0.03$ \\
$\pm 1 \sigma$:  &   0.18   &  0.16  &   0.22  &  0.20  & 0.19 \\

\enddata
\tablecomments{The first three columns
  present the mean evolution in $M/L_V$ broken down by morphological
  type, both with and without galaxies of $\sigma < 100$ km
  s$^{-1}$, and for the whole cluster sample or just galaxies with
  projected radius $R<1$~Mpc. Column four includes all our E+S0s, but excludes two
  disturbed-morphology galaxies with outlying values of
  $\Delta \log \left( M/L_V \right)$. In the fifth column, we add
  our data to the eight galaxies from \citet{vdf} that do not overlap
  with our own. $\pm1\sigma$ values are logarithmic, representing the
  scatter in  $\Delta\log \left( M/L_V \right)$. }
\end{deluxetable*}

Open squares in Figure~\ref{fig:fp}
indicate points from \citet{vdf}, who also measured the FP in Cl~0024.
The two sets of data fall on the same
plane, and our inferred $\left< \Delta \log \left( M/L_V \right) \right>$ is
consistent with their work: they calculated $\left< \Delta \log \left(
M/L_V \right) \right>=-0.12\pm0.03$  Our much larger sample should allow for
greater precision in calculation of the FP zero point, yet the two
measurements yield similar uncertainties in $\left<
\Delta \log \left( M/L_V \right) \right>$. This is due to the
surprisingly high scatter that we uncover in the Cl~0024 FP, which was
not apparent in the earlier sample of 9 galaxies in \citet{vdf}.

Importantly,  we find the intrinsic scatter in the FP of Cl~0024 is  $40\%$ in
$M/L_V$, significantly higher than the $\sim20\%$ found locally
\citep[e.g.][]{jfk96}, and also higher than the $\sim13\%$ scatter in
the FP of Cl~1358+62 at $z=0.33$ \citep{kelson00c}. While we probe
the FP for a larger range of galaxy masses than \citet{kelson00c},
we see a higher scatter even in the more massive galaxies that are
comparable to those in their study. Interestingly, this increased
scatter is due almost entirely to an enhanced scatter in $M/L_V$
for galaxies within the inner 1~Mpc of the cluster, a region that
has been well studied in other clusters in this redshift range
\citep{bender98,kelson00c, ziegler01}. 
To verify this increased scatter is a true physical effect, 
we examined the possibility that errors in our surface photometry are 
higher in the more crowded cluster core, but found that galaxies with nearby
neighbors cannot account for the high scatter. We also checked for
errors in our velocity dispersion measurements: when removing five
early type spectra ($\sigma>100$ km~s$^{-1}$) with $S/N < 10$ from 
our high quality sample, we find no significant change in our FP
zero point or scatter.

A similarly large
scatter was found by \citet{wuyts04} for the cluster MS 2053-04 at
$z=0.58$ ($\sim42\%$), so this may be an effect seen
only in some fraction of intermediate redshift clusters. At even
higher redshift, \citet{holden05} finds a large scatter in $M/L_B$ for
massive early types at $<z>=1.25$.  We will
defer further discussion of this effect to \S~4.2, where we
discuss radial trends.

Recent results indicate that the parameters of the FP at intermediate 
redshift may differ from the local values 
\citep[e.g. Treu et al. 2005a,b,][]{wuyts04,vanderwel05}. 
We investigated using the method of \citet{jfk96} to derive an 
independent fit for the parameters of the FP from our sample, 
including Montecarlo simulations to account for bias in our 
magnitude-limited sample, but found no conclusive evidence for 
a change in FP parameters from the local values. 

Nevertheless, the group of galaxies with $\sigma<100$ km s$^{-1}$ in
Figure~\ref{fig:fp}, located at the
lower left of the plot, seem to deviate significantly from the
FP. As most calculations of the FP parameters specifically exclude
galaxies with $\sigma < 100$ km s$^{-1}$ \citep[e.g.][]{jfk96,
  lucey91}, including our own, it is not surprising that
such galaxies in our sample deviate from the FP.  Yet there is also
evidence that, in the local universe, these small ellipticals and
dwarf ellipticals may behave differently as a population than
larger E+S0s \citep[e.g.][]{burstein84}. Our data suggest that this population split may
have existed already at $z\sim0.4$.
S
When dividing our sample into three subclasses (E, S0, and
E/S0), we find no evidence of significant variation between the populations,
either in scatter or intercept. Table~4 lists values of $\left< \Delta
\log \left( M/L_V \right) \right>$ and the scatter about the mean for
several different subsamples of our data. For galaxies with
$\sigma>100$ km s$^{-1}$ in particular, there appears to be no
variation at all between E, E/S0, and S0 morphological types. This
lack of variation is in disagreement with recent
results by \citet{fritz05}, who marginally detect a difference between
ellipticals and S0s in two clusters at $z\sim0.2$, equivalent to
a difference in $\delta\left< Log \left( M/L_V \right) \right>$ of
$\sim-0.16\pm0.11$ (with S0s being the younger/more luminous
population). 
On the other hand, \citet{kelson01} find no such
distinction between ellipticals and S0s at $z=0.33$. Our sample is
larger than either of these other studies, and so we place stronger
constraints on the possible $M/L_V$ variations between Es and S0s at
$z\sim0.4$.  As discussed in $\S$2, we attempted to improve our E:S0 classification
by examining residuals from our surface photometry fits as a means
of locating faint disks or bars (See Figure~\ref{fig:galfits}). Two of us
(RSE and TT) independently examined the residuals and, although
the revised classes agreed fairly closely, no significant FP differences 
between ellipticals and S0s were found. 

\subsubsection{$[MgFe]^\prime$--$\sigma$ and Balmer--$\sigma$ relations}

While the FP helps to connect the dynamic and photometric structure of
an early type galaxy to the overall mass to light ratio of its stellar
population, we can seek an improved picture of a galaxy's stellar
population by also examining correlations between spectral line strengths
and $\sigma$.  Local early type galaxies exhibit a tight correlation
between the $Mg_2$ line index and $\sigma$, for example
\citep[e.g.][]{kunt01}. The narrower
$Mg^b$ index is centered over the same spectral feature, and, at least
locally, behaves in the same way as $Mg_2$\citep{burstein84}. Similarly,
correlations are observed between $\sigma$ and the strengths of
several Balmer lines: H$\beta$, H$\gamma$ and H$\delta$ \citep{kelson01, kunt01}.

Our current set of high quality spectra of early types in Cl~0024
provides a large sample of galaxies at $z\sim0.4$ where diagnostic
spectral lines can be measured with some accuracy for the first time. We measure
$Mg^b$, $Mg_2$ and the composite index $[MgFe]^\prime$ to make a broad
comparison between the metallicity features of early types at intermediate
redshift and those seen locally.

We also measure Balmer line strengths, which are commonly used to
judge the age of early type galaxies \citep{worthey97}.
However, low levels of recent star formation can make an early-type galaxy's
stellar population appear much younger than the true average age of
its stars. For the first 1 Gyr after a starburst, the strong Balmer absorption
present in the spectra of A stars can be prominent in the integrated
galaxy spectrum. The H$\beta$ index is a particularly age-sensitive indicator,
but is affected by contamination from emission. While H$\alpha$ can be used to correct
H$\beta$ for small amounts of emission, this line falls outside of the
range of our spectra, so we choose to study the H$\delta$ and
H$\gamma$ lines instead, which are relatively unaffected by emission.
Measuring H$\delta$ and H$\gamma$ in Cl~0024 can thus
allow us to probe both galaxy age and episodes of recent star formation,
if the two effects can be separated.

Early attempts to trace the evolution of these relations with redshift
has produced a variety of results. \citet{kelson01} reported a 
correlation between (H$\delta+$H$\gamma$)
and $\sigma$ for several clusters between redshifts of 0.08 and
0.83. The slope they measure for Cl 1358+62 at $z=0.33$ is consistent
with their data at all other redshifts, though they are unable to
measure slope for each cluster individually. They plot the zero point
evolution of this relation, and find that the change in mean Balmer
absorption strength from $z=0.83$ to the present epoch is consistent
with passive evolution models of early types, and constrain $z_f>2.4$.
Likewise, \citet{barr05} study a poor cluster at $z=0.28$ and find
that the zero point evolution of the (H$\delta+$H$\gamma$)--$\sigma$ 
relation, when compared to \citet{kelson01}, is only marginally
inconsistent with passive evolution.

Conversely, the evolution of the metallicity correlations has
proven more complicated. 
The question of using spectral line strengths to uniquely determine ages,
metallicities, and $\alpha$-element enhancement ratios for elliptical 
galaxies is not
yet solved even for local galaxies, as made clear by the excellent
study of the problem by \citet{tantalo04}. Further, our data likely
fall short of the $S/N$ needed to determine these galaxy properties
reliably \citep{trager98}. It is therefore beyond the
scope of this paper to examine how the
various Mg--$\sigma$, $[MgFe]^\prime$--$\sigma$ and Balmer--$\sigma$
relations evolve with redshift.

However, we note that correlations with velocity dispersion do
exist as far back as $z\sim0.4$, regardless of the origin of such
a relation. Secondly, large deviations from these relations may
indicate a recent episode of star formation in that galaxy. It
will be useful then to look later at the residuals from the
$[MgFe]^\prime$ and Balmer--$\sigma$ relations, as a function of
cluster environment.

\begin{figure}[t]
\centering
\includegraphics[width=\columnwidth]{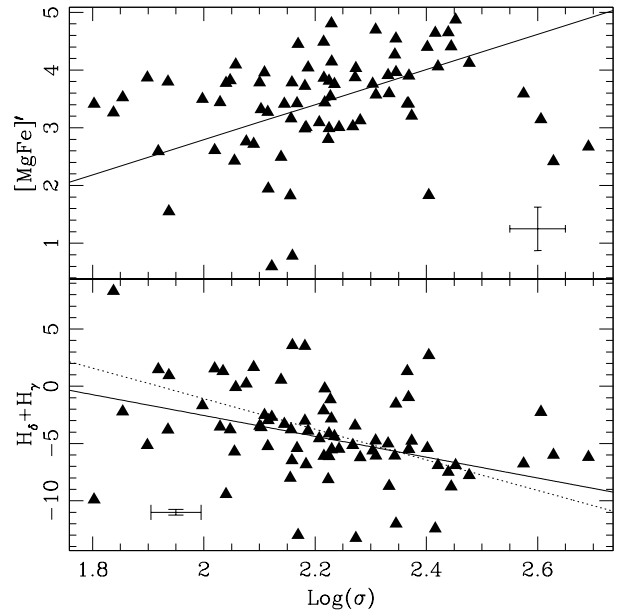}
\caption{\label{fig:mgfe_sig}Top: $[MgFe]^\prime$--$\sigma$ at resolution
  of 3$\mbox{\AA}$. The solid line is the best least-squares fit to
  our data, excluding the outlier points at $\sigma \ge 300$ km
  s$^{-1}$.
Bottom: $(H\gamma+H\delta)$--$\sigma$ at
resolution of $\sim10\mbox{\AA}$, compared to \citet{kelson01}, at same
spectral resolution. Indices are corrected to match the aperture
  used by Kelson et al. The solid line represents the best-fit relation
from \citet{kelson01}. We adopt the slope of their relation, but
calculate the intercept at $z=0.4$ by interpolating between their data
points at $z=0.33$ and $z=0.58$. The dashed line is the line of best fit to our
data. The scatter is large, so our best-fit relation is highly
  uncertain. In both panels, velocity dispersions are aperture
  corrected to a $3\farcs4$ diameter aperture at the distance of the
  Coma cluster.}
\end{figure}

We can minimize uncertainties due to variations in abundance
ratio by focusing on the $[MgFe]^\prime$ index, which should provide a
reliable estimate of the total metallicity of a galaxy \citep{thomas03}.
Figure~\ref{fig:mgfe_sig} presents the correlation between
$[MgFe]^\prime$ and $\sigma$. The bulk of galaxies measured fall along
the solid best-fit line plotted, though the scatter is high. The high
scatter could be due to the large error bars on each measurement.
However, there are a significant number of outliers with anomalously
low $[MgFe]^\prime$ that seem not to be due to measurement errors.
These outliers correlate with other properties of the each galaxy's stellar
population, and will be discussed further in \S~5.

Figure~\ref{fig:mgfe_sig} presents (H$\delta+$H$\gamma$)
vs. $\sigma$ for Cl~0024. The two quantities correlate in the sense
that the largest, highest $\sigma$ galaxies exhibit the weakest Balmer
absorption. This is as expected if the stellar populations are oldest
in the largest elliptical galaxies. Overplotted in
Figure~\ref{fig:mgfe_sig} is the best-fit relation from
\citet{kelson01}, where we adopt their slope and interpolate between
their points at $z=0.33$ and $z=0.58$ to determine the zero point at
$z=0.39$. Our data is consistent with their findings, though again our
scatter is quite large. We also overplot our best fit relation as a
dotted line, though the high scatter makes this relation highly uncertain.

In fact, just as we saw with the FP of Cl~0024, we see a scatter that
is much larger than that observed in Cl~1358+62 at similar
redshift. Both the FP and (H$\delta+$H$\gamma$)--$\sigma$ relation are
reported to be much tighter for Cl~1358+62 than for Cl~0024
\citep{kelson01, kelson00c}. On the other hand, \citet{barr05} also 
find an increased scatter in the (H$\delta+$H$\gamma$)--$\sigma$
relation at $z=0.28$, though they do not measure the FP. 
As with the FP and Metallicity--$\sigma$ relation in Cl~0024, some of the
enhanced scatter we observe in Balmer line strengths is a direct 
signature of environmental evolution, and will be discussed in \S~4.2.3 below.

\subsection{Radial Trends}

We now turn to discuss environmental trends present in the early type
population of Cl~0024. For the brighter sample of galaxies with high
quality spectra, we can analyze the residuals from the empirical
scaling laws presented in \S~4, and look for variations with local
density, cluster radius, or galaxy luminosity. For the full sample of
galaxies to $I=22.5$ we additionally examine the environmental
variations in several key spectral lines.

Paper I in this series found that local density was the primary
measure of environment outside of the cluster core, suggesting
that infalling groups are the logical unit out of which clusters
are built. Looking at the detailed spectral properties of early-types 
now, it
would be useful to determine whether spectral properties depend
more closely on local density, as the morphological mix seems to,
or on radius. Because we are only tracing a much smaller
population of 104 galaxies of a single morphological class, we are
unable to make such a distinction reliably. We will, however,
present plots of spectral features as a function of both R and
$\Sigma_{10}$. Because projected radius R is more easily measured
than $\Sigma_{10}$, we will focus our discussion on radial trends.
Anticipating the results, we note that the observed trends with R
and $\Sigma_{10}$ are qualitatively similar.

\subsubsection{Fundamental Plane}

\begin{figure}
\centering
\includegraphics*[width=1.05\columnwidth]{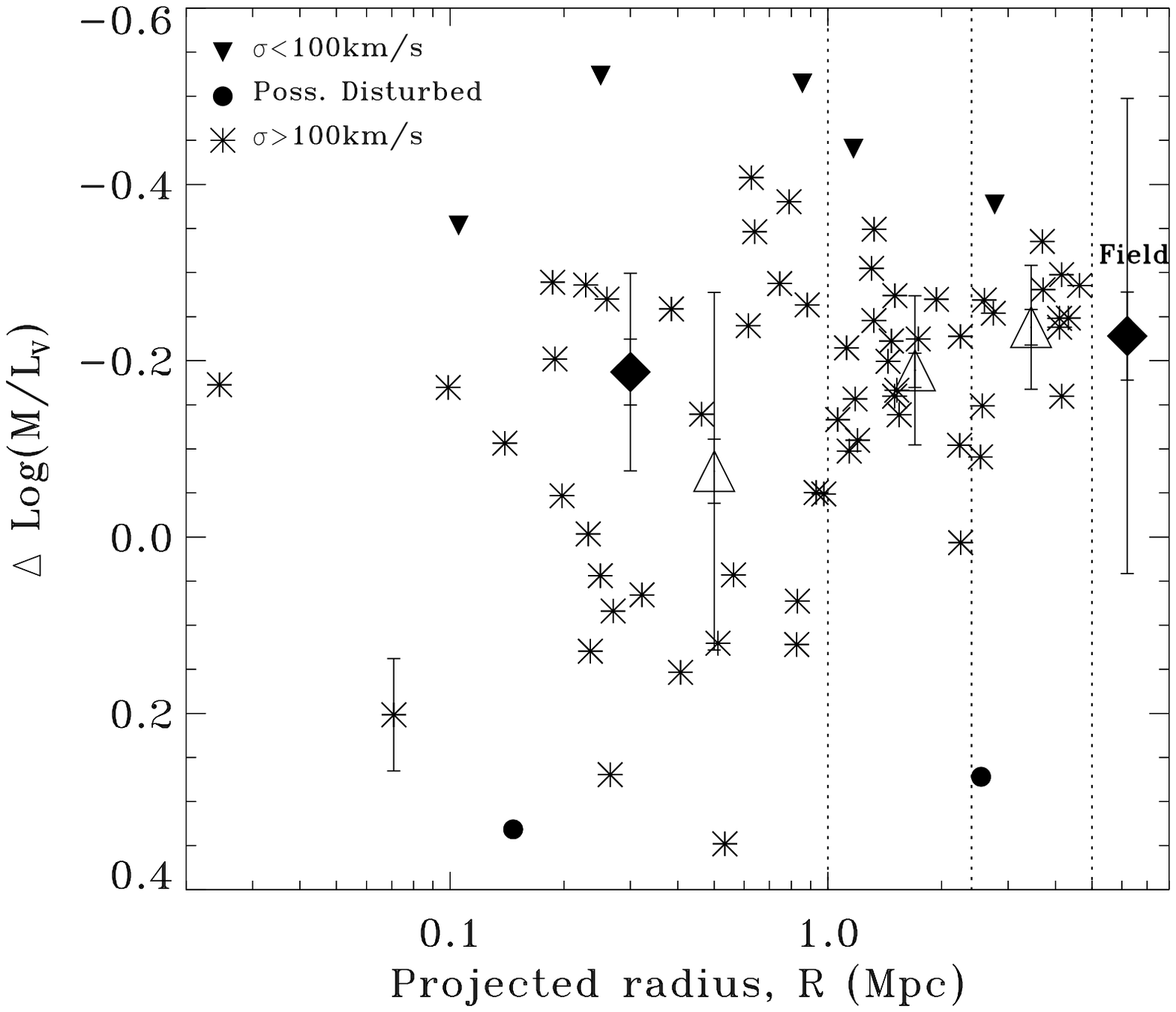}
\includegraphics*[width=1.05\columnwidth]{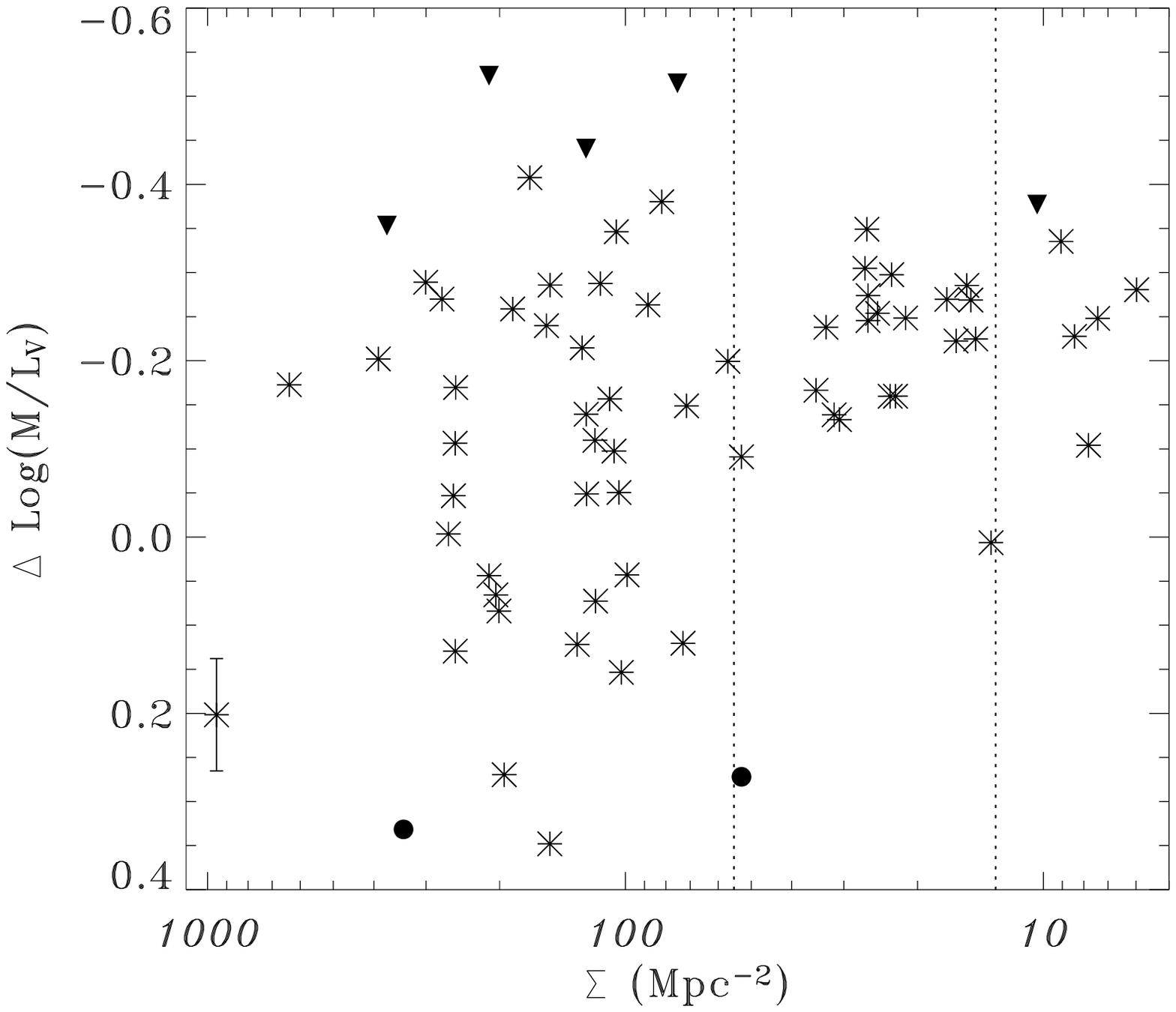}
  \caption{\label{fig:deltaml}Change in $M/L_V$ ratio for galaxies in Cl~0024,
with respect to $\left< M/L_V \right>$ for Coma. Plotted with
respect to radius, top, and local density, bottom. Asterisks
represent individual galaxies, with a typical error bar shown on a
point near the lower left of each plot. For the upper plot, open
triangles represent averages for the three radial zones indicated.
Large error bars reflect the rms scatter, and small bars are the
error on the mean. Filled
 diamonds are points from \citet{vdf} and \citet{tt05b} for cluster
core and field, respectively. Dotted lines at 1~Mpc and 2.4~Mpc delimit the three radial
zones discussed in the text; corresponding dotted lines in the local
density plot represent the mean values of $\Sigma_{10}$ at 1~Mpc and
2.4~Mpc, respectively.}
\end{figure}

The high scatter about the FP of Cl~0024, discussed in \S~4.1
above, leads us to examine $\Delta Log \left( M/L_V \right)$ on a
galaxy by galaxy basis.  For an individual early type (labeled by
subscript {\it i}), the evolution in $M/L_V$ with respect to the
prediction of the local FP is related to its offset from the local
intercept $\gamma$ according to:
\\
\begin{displaymath}
\Delta\gamma^i =\log R_e^i-\alpha\log\sigma^i-\beta\mu_V^i-\gamma
\end{displaymath}
and 
\begin{displaymath}
\Delta\log\left( \frac{M}{L_V} \right)^i = -\frac{\Delta\gamma^i}{2.5\beta}
\end{displaymath}
\\
In Figure~\ref{fig:deltaml}, we plot
$\Delta Log \left( M/L_V \right)$ as a function of projected
radius and local density. Both plots show a trend toward increased
scatter near the cluster core, at small radius and high densities.

In the radial plot, the open triangles overplotted represent the
mean $\Delta Log \left( M/L_V \right)$ for galaxies within each of
the three radial zones delimited by dotted lines: cluster core ($R
< 1$ Mpc), transition region ($ 1 \le R < 2.4$ Mpc), and periphery
($R \ge 2.4$ Mpc). These three zones were introduced in Paper I,
and represent regimes where different physical mechanisms may be
effective in transforming galaxies. We see a clear trend of
increasingly negative $\Delta Log \left( M/L_V \right)$ as we move
outward, indicating that galaxies are more luminous for their
mass, and hence appear younger, at larger distances from the
cluster center.

Galaxies marked with a filled triangle in Figure~\ref{fig:deltaml}
have $\sigma < 100$ km s$^{-1}$. These, along with two
disturbed-morphology galaxies marked with filled circles, are not
included in the calculation of the means for each radial zone. The
low--$\sigma$ galaxies are likely biased toward low $M/L_V$, due
to the selection effect where only the brightest of the small, compact
cluster members are within our limiting magnitude.

The solid diamond plotted within 1 Mpc on Figure~\ref{fig:deltaml}
is the mean $\Delta Log (M/L_V)$ found by \citet{vdf}; it is clear
now why our mean value of $\Delta Log (M/L_V)$ is slightly larger
than theirs: the additional contribution from lower $M/L_V$
galaxies in the transition region and periphery boost the mean
luminosity evolution we calculate. It is interesting also to note
that our new wide-field data bridges the gap between the mass to
light ratios typical of cluster cores and those found in the field
at this redshift. The other solid diamond plotted in
Figure~\ref{fig:deltaml}, labeled ``Field'', is from
\citet{tt05b}, who measured the FP of field galaxies at
intermediate redshift. Similar values for the field are reported by
\citet{tt01} and \citet{vanderwel05}. While some studies of the field
FP at intermediate redshift have reported a slower redshift evolution 
\citep[e.g.][]{ziegler05, vdf01, rusin03}, \citet{vanderwel05} have 
shown that most of the discrepancy between studies of field early 
types can be attributed to differences in analysis methods and 
selection criteria. Since the reported variations are small on the scale of 
Figure~\ref{fig:deltaml} (less than $\pm0.1$ in $\log(M/L_V)$), 
we can be confident that $\left< M/L_V \right>$ does actually vary with
environment at intermediate redshift.  This gradient in 
$\left< M/L_V \right>$ across the cluster appears to directly reflect 
the difference in formation history between galaxies in high and low density
environments. 

Within the cluster core, the scatter in $\Delta Log (M/L_V)$ is
highest, as noted above in \S~4. We caution that projection
effects could contribute to the apparent scatter near the cluster
center, especially if there is a strong gradient in $M/L_V$ with
3D radius, r. However, for any sensible distribution, we would 
expect that the number of true core members is  
higher than those at large radii projected upon the core. 
As we see approximately equal numbers of galaxies both above and below
the mean $\Delta Log \left(M/L_V \right)$ in the core, we conclude
that projection effects cannot account for the increased
scatter. This is confirmed by Kolmogorov--Smirnov tests that compare 
various simulated distributions of galaxies to our observed distribution.

Similarly, we find no evidence that the high- or low-$M/L_V$ early
types in the core are part of an infalling group; $\Delta Log
\left(M/L_V \right)$ appears uncorrelated with the velocities or 
spatial distribution. It has been hypothesized that Cl~0024 is
currently undergoing a face-on merger with a large group, as indicated
by the double-peaked redshift distribution first reported
by \citet{czoske}, and replicated in the redshift distribution of
Figure~\ref{fig:redshifts}. However, none of the galaxies on our FP
lie in the secondary ``Peak B''.
While the high scatter may still be related to the subcluster merger, 
whatever mechanism affects the mass to light ratios of early-types 
in the cluster core is not apparent in any other measurements.

Although the observed scatter seems at variance with
the notion that massive cluster ellipticals are uniformly old and quiescent,
if we use the \citet{bc03} SSP models to predict the passive evolution
of these galaxies to $z=0$, the
resulting scatter will match that observed locally. Furthermore, the scatter in the field FP reported by
\citet{tt05b} in this redshift range is larger than in the core of
Cl~0024, as expected if the range
in ages and star formation histories is greater in the field than in
the cluster environment.

There may be two separate populations of E+S0s in the
core, then: older galaxies that formed earliest and which have
$M/L_V$ already similar to that of local E+S0s, and galaxies with
a lower $M/L_V$ which have more recently fallen into the cluster
core. While there is no obvious separation between these two
groups in Figure~\ref{fig:deltaml}, we do notice a residual
correlation between $\Delta Log (M/L_V)$ and $\sigma$, in the
sense that the most massive, highest--$\sigma$ cluster galaxies
also have the highest mass to light ratios, indicating the oldest
stellar populations. This is not a surprising correlation, but it
does confirm that the high $M/L_V$ galaxies are typically the
oldest and largest found in the cluster. It is also in agreement
with a more general trend identified by \citet{tt05}, who found
that less massive galaxies exhibit a younger luminosity-weighted
age than do the most massive ones. 

It is important to note, however, that the radial trend observed in
$\Delta Log \left(M/L_V \right)$ is not due solely to this ``downsizing''
relation between galaxy mass and $M/L_V$. There is still a radial
gradient in $M/L_V$, even when we further restrict our sample to a narrow
range of galaxy masses\footnote{We calculate a galaxy's dynamical
mass, in solar units, according to $M=5\sigma^2 R_e/G$, which is
equal to $M=1162 \sigma_0^2 R_e$, with
$R_e$ in pc and $\sigma_0$ in km~s$^{-1}$.}. For example, if we limit our sample to
the mass range $10.9 < Log (M/M_\odot) \le 11.4$, we
observe a significant difference in $M/L_V$ between 
the cluster periphery and core: $\delta<Log (M/L_V)>=0.13\pm0.07$.
Using field galaxies from the \citet{tt05b} sample within this same mass
range (and in the redshift range $0.3 < z < 0.5$), we find an offset
between the Cl~0024 cluster core and the field equal to 
$\delta<Log (M/L_V)>=0.09 \pm0.06$.
The lower mass limit was chosen because
galaxies above $Log (M/M_\odot)=10.9$ are negligibly biased due to
the luminosity limit of both surveys. The upper mass limit is 
chosen so that we include only galaxies that are
well-represented across the entire cluster, excluding the
high-mass galaxies found only in the core.

It makes sense to search for morphological differences between the old
and young E+S0s in the cluster core. The existence of the
morphology--density relation in Cl~0024 (Paper I) leads us to consider
whether the younger, low $M/L_V$ galaxies are preferentially S0s
that have recently been created or transformed from spirals. However, when we
divide our sample into groups that lie above and below the mean
$M/L_V$ in the cluster core, we see no evidence for
morphological segregation. 

In the second paper in this series on Cl~0024, \citet[][hereafter,
Paper II]{kneib03} estimated the global mass to light ratio of the
cluster as a function of radius, by combining a mass map based on their
gravitational lensing analysis with $K_S$-band observations which
trace the stellar mass. They concluded that the overall $M/L$ of the
cluster is remarkably flat, at least within 2 Mpc of the cluster
core.(Their result was the same whether they measured total light via
K or rest-frame V observations.) At first glance, the radial gradient
in $M/L_V$ seen in Figure~\ref{fig:deltaml} may appear to contradict
the results of Paper II. However, the {\it cluster} $M/L$ ratio measured in Paper II is much more strongly
dominated by dark matter than the {\it early type galaxy} $M/L_V$ discussed here, and
so variations in mass to light ratios of individual galaxies can be
consistent with an overall cluster $M/L_V$ that is constant.

\subsubsection{Line Strengths}

In the next two sub-sections, we will consider environmental trends in
several different indicators of star formation and metallicity, in both our
high-quality sample of galaxies and the full sample to $I=22.5$. It
is important to gauge the effect of any recent star formation on the
overall $M/L_V$ of these galaxies, to ensure that we can correctly
attribute differences in $M/L_V$ to differences in mean galaxy age,
rather than the effects of relatively recent and small star forming
events. We also wish to trace the star formation itself, as any
significant star formation in early-type galaxies can act as a
signpost to the different physical mechanisms that could be at work.

\begin{figure}[t]
\centering
\includegraphics[width=\columnwidth]{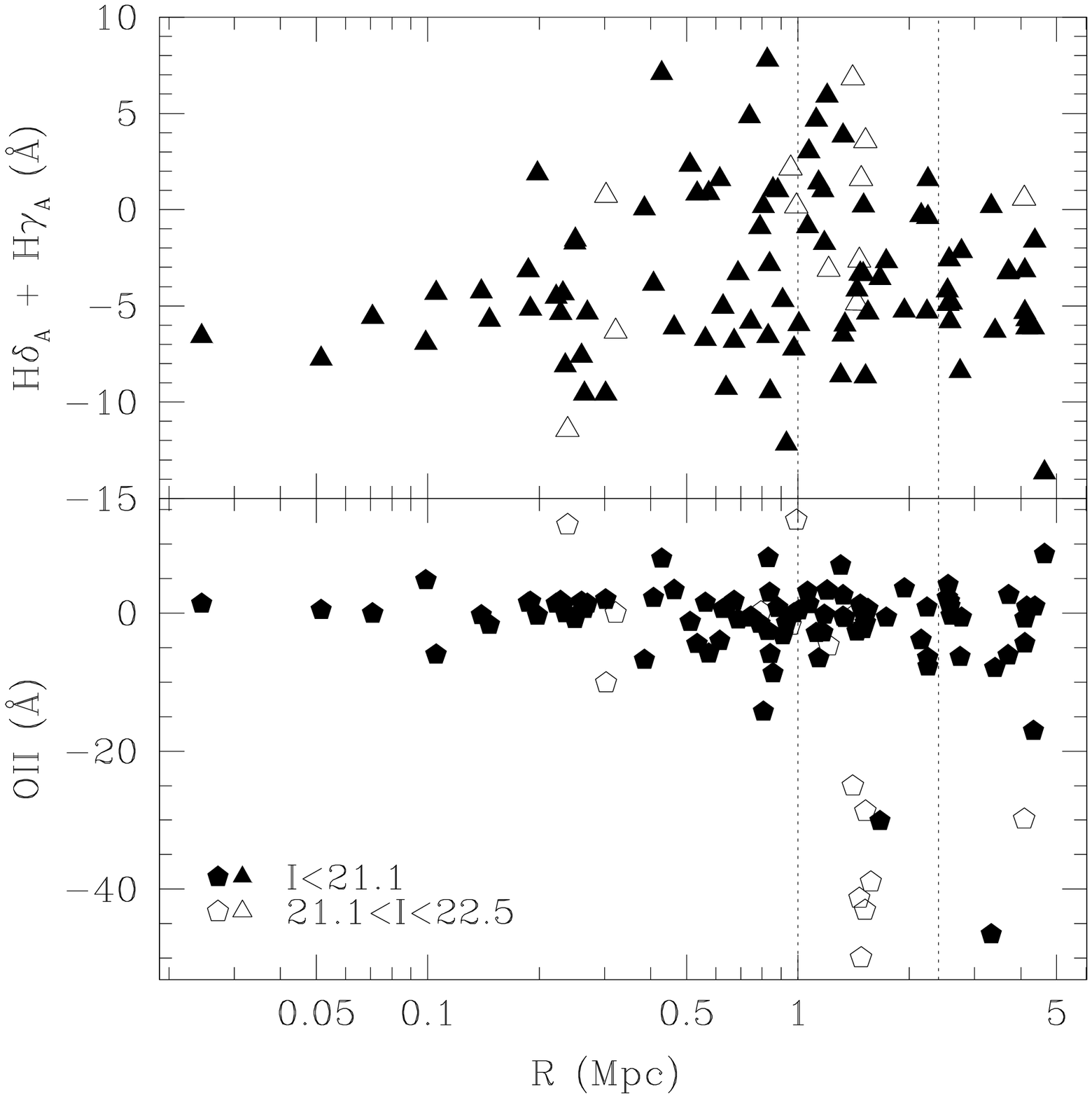}

\includegraphics[width=\columnwidth]{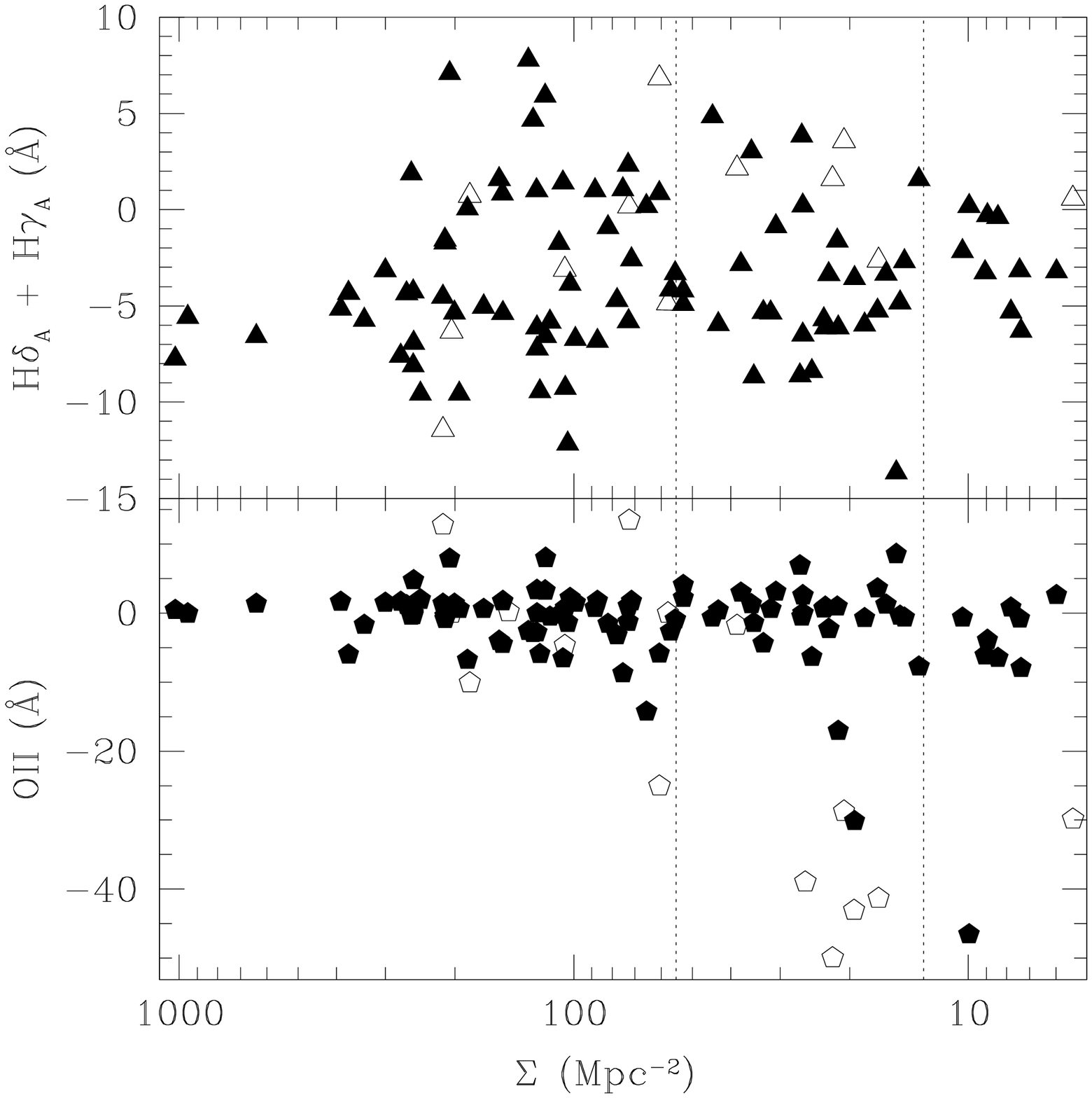}
\caption{\label{fig:ew}H$\delta_A +$H$\gamma_A$ and [OII] EWs for all
  E+S0 members as function of radius (top) and local density
  (bottom). Negative values represent emission in [OII], but can be
  considered weak absorption for the Balmer lines, due to an effect
  where the continuum flux in the index sidebands falls completely below the
  flux in the index bandpass. In the upper plot, dotted
  lines delineate the core, transition region, and periphery. Dotted
  lines in the lower plot mark the mean values of the local density at
  these same radii. Open symbols: $21.1<I<22.5$. Filled symbols: $I<21.1$.}
\end{figure}

We focus on two key indicators of star formation, the [OII] emission
line, and the combination (H$\delta+$H$\gamma$), which respectively
measure ongoing star formation, and recently completed star
formation, as discussed in \S~3. Figure~\ref{fig:ew} plots the
equivalent widths of these two diagnostic indices as a
function of radius and local density. Solid symbols indicate galaxies
with $I\le21.1$; we note that some of these brighter galaxies do not
have measured velocity dispersions, and so were not included
in the high quality sample discussed in previous sections. 
Open symbols are galaxies in the magnitude range
$21.1<I\le22.5$. As we saw with the FP, we note that the
strengths of these indices vary in a similar way with both local
density and projected radius.

\begin{figure*}
\plottwo{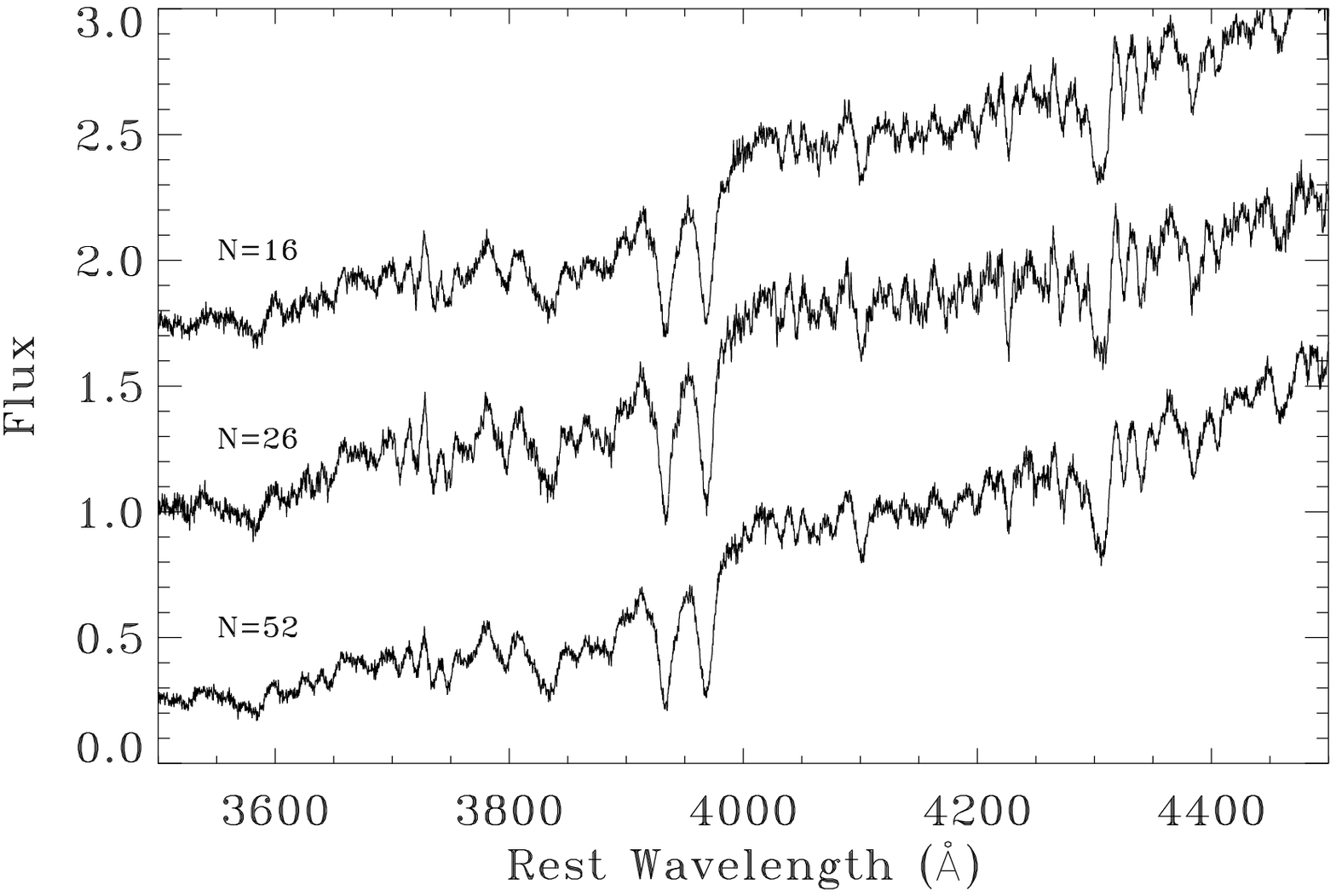}{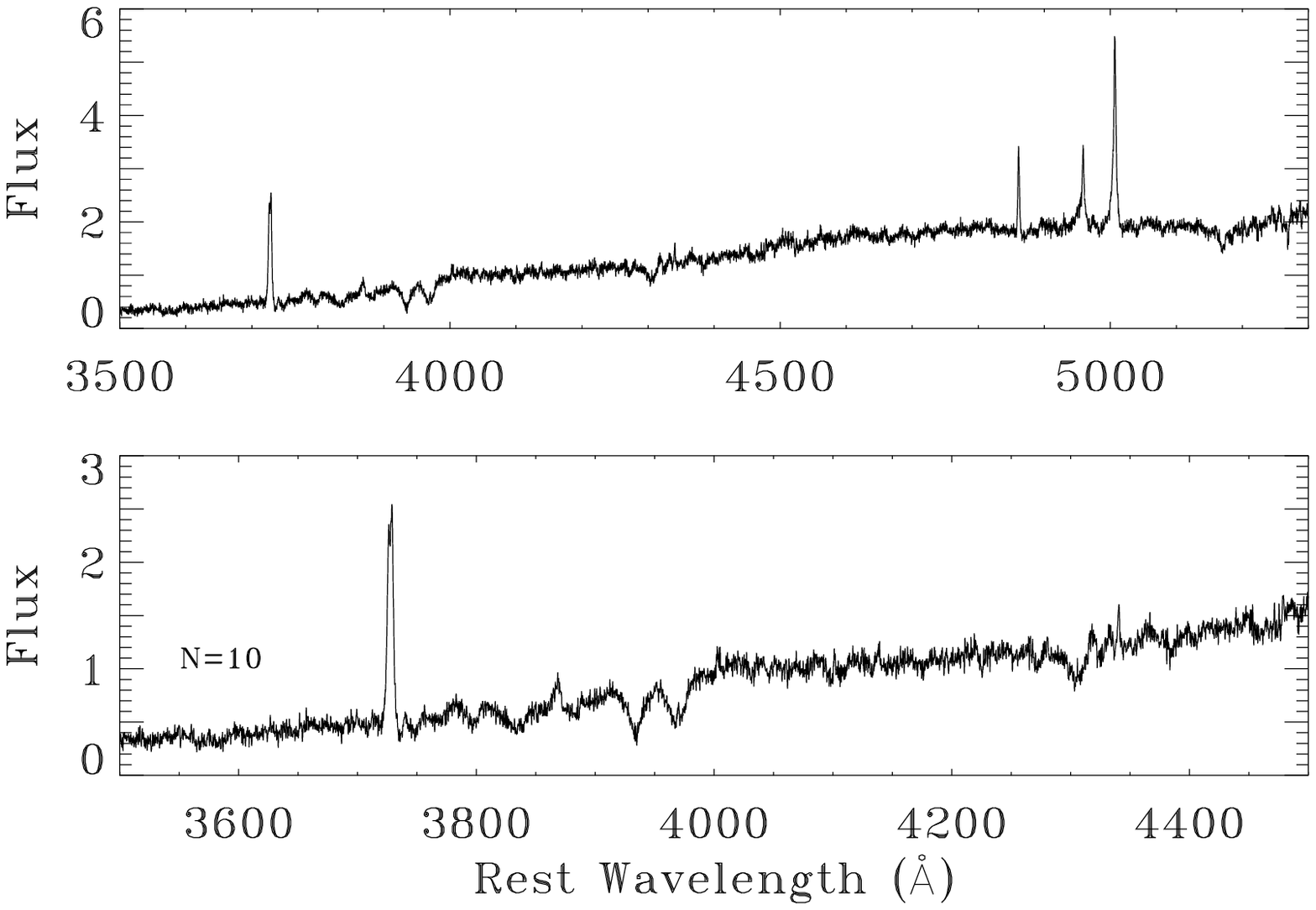}
\caption{\label{fig:coadded} Coadded spectra.
  The right panel shows coadded spectrum of E+S0 members
  with EW(OII) $< -15 \mbox{\AA}$. The full spectrum is on top, with
  an excerpt below it which shows OII emission, Ca H \& K, and Balmer
  absorption lines in more detail. The left panel shows coadded
  spectra for all other E+S0 members, divided by distance from
  cluster core. From top to bottom: periphery, transition region,
  and cluster core. Wavelengths in rest frame.}
\end{figure*}

Examining individual measurements of EW([OII]), we find a
population of galaxies with strong emission, which are
concentrated in a narrow range in radius, close to the Virial
radius at 1.7 Mpc. The spectra of these galaxies are dominated by
emission lines, including [OIII] and several Balmer lines. (See
Figure~\ref{fig:coadded}). These emission line galaxies are
preferentially dim; as denoted by the un-filled pentagons in
Figure~\ref{fig:ew}, most are in the magnitude range $21.1 < I \le
22.5$. Most previous studies have not been sensitive to early-type
cluster galaxies at these luminosities and in this radius range, though
a recent survey of a cluster at $z=0.83$ by \citet{homeier} has
uncovered a similar population of dim emission-line E+S0s, discussed
in more detail in \S~5.

\begin{figure}[t]
\centering
\includegraphics[width=\columnwidth]{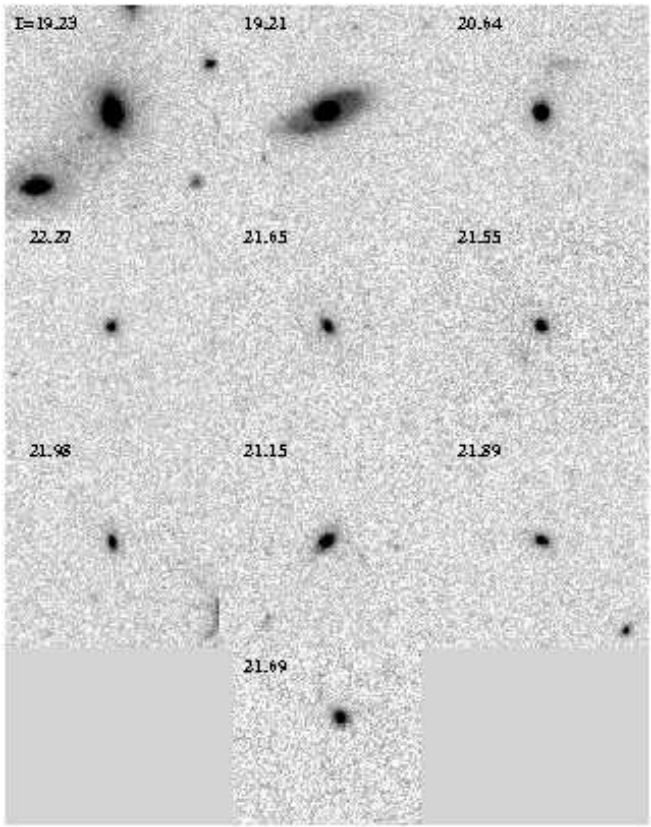}
\caption{\label{fig:oiimontage} Postage stamp images of all galaxies
  with EW(OII) $< -15 \mbox{\AA}$. The top row shows galaxies with
  $I \le 21.1$, and the lower rows contain galaxies in the range $21.1 < I
  \le 22.5$. Stamps are 10\arcsec on a side.}
\end{figure}

What could be the nature of these emission
line galaxies clustered in radius? While we leave most of the
discussion of this question to \S~5, we can quickly address some of the
possibilities. Figure~\ref{fig:oiimontage} shows
postage stamp images of each of these active galaxies. The top row
contains the three emission line galaxies with $I<21.1$; one of
these may be interacting with a neighbor, though the redshift of
the neighboring galaxy is not known. None of the others seem to 
be undergoing major mergers, so merger-induced star formation is 
ruled out as a cause of emission. One other bright
[OII]-emitter could possibly be a misclassified spiral. We looked at
the residuals from surface photometry fits for the rest of these
galaxies, but found no other indication of spiral arms. 
Seven of these ten galaxies are also detected by the $H\alpha$ imaging 
survey of \citet{kodama04}, and two of the three non-detections have
redshifts that would likely place the $H\alpha$ line outside the
bandwidth of the narrow-band filter employed by Kodama et al.

Line ratio tests meant to distinguish between star formation and
AGN activity are inconclusive for these galaxies. XMM-Newton
observations are available which cover a field including most of
these emission line early types: none are associated with an X-ray
point source \citep{zhang05}, even though
the XMM-Newton observations are deep enough to detect any
clear-cut AGN with $L_X>10^{42}$ erg~s$^{-1}$. Therefore, while we
cannot rule out AGN activity for these galaxies, none are
definitively AGN. An interesting hypothesis
that we will explore in \S~5 is that these galaxies have
suffered harassment or disturbances from their
interaction with the intra-cluster medium (ICM). 

In addition to the emission line galaxies, we see a higher mean 
EW([OII]) at larger
radius. Figure~\ref{fig:coadded} shows the coadded normalized spectra of all
early-types
binned by radial zone; top to bottom, they trace the mean early-type
spectrum from the periphery to the core.
All galaxies with EW(OII) $< -15 \mbox{\AA}$ were coadded
separately; see Figure~\ref{fig:coadded}, right.  Co-added spectra of
`normal' E+S0 members in Figure~\ref{fig:coadded} show an increase in an
average galaxy's OII emission outside of the cluster core.
This trend is qualitatively similar to radial gradients in emission
line strength seen for samples of cluster galaxies that 
include all morphologies, both at intermediate
\citep[e.g.][]{gerken04} and low redshift \citep[e.g.]{gomez03, lewis02}.

\begin{figure}
\centering
\includegraphics[width=\columnwidth]{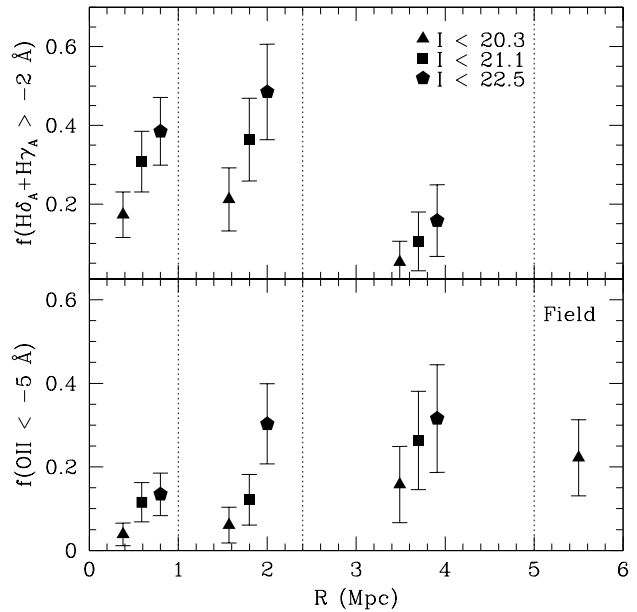}
\caption{\label{fig:OIIfrac}
  Fraction of E+S0 galaxies with EW(H$\delta_A+$H$\gamma_A) >
  -2\mbox{\AA}$, top, and EW([OII]) $< -5\mbox{\AA}$,
  bottom, as a function of radial zone. Triangles represent
  fraction for galaxies with $I<20.3$. The field value is from
  \citet{tt02}. Squares and pentagons are the same
  measure, but to $I=21.1$ and $I=22.5$, respectively.}
\end{figure}

In Figure~\ref{fig:OIIfrac}, we plot the fraction of all E+S0
galaxies with EW([OII])$< -5 \mbox{\AA}$, averaged across each of
the radial zones indicated. At all luminosities, it is clear that
the fraction of [OII] emitting early types is highest in the
cluster periphery and field. Though the fraction of emission line galaxies
may be elevated due to selection effects (\S~3), especially at
fainter magnitudes, there still exists radial variation within any
single magnitude bin. This radial variation holds (in the two brighter
magnitude bins) even if we 
exclude the cluster core, where the
fraction of [OII] emitters may be affected by the process causing the
strong [OII] emission at the Virial radius.

We see then that both the fraction of
galaxies with measurable [OII] emission, and the average strength
of that emission rises slightly with radius. This suggests an
encounter with the cluster environment which serves to gradually suppress
the already low levels of early-type star formation during infall.
Another possibility is a
simple gradient in galaxy formation age: older early types near
the cluster core may simply contain less residual gas available
for star formation. However, this scenario is more difficult to
reconcile with other studies that have found strong gradients in
the overall star formation rates within clusters 
\citep[e.g.][]{kodama04,pogg99,balogh00}.

Turning now to the Balmer absorption strengths, we see a mix of
galaxies in Figure~\ref{fig:ew} with low to moderate
Balmer absorption in the periphery and transition region. These are
consistent with the low levels of ongoing star formation indicated by [OII]
emission. Almost coincident with the set of strong [OII]-emitters
discussed above, we begin to see galaxies with much higher values of
(H$\delta$+H$\gamma$), which appears to decay toward the center of the
cluster. If the [OII] emission seen at the Virial radius is, in fact,
due to star formation, these galaxies with enhanced Balmer absorption
may have undergone a similar burst of star formation in the recent
past. The excess absorption then seems to decay away as galaxies migrate
further toward the center of the cluster.

Figure~\ref{fig:OIIfrac} confirms that the fraction of early-types with
strong Balmer absorption rises dramatically for galaxies in the
transition region, and stays high even in the cluster core. Since so
few early types in the periphery show similarly high Balmer
absorption, there is likely an interaction with the cluster
environment that triggers both the enhanced OII emission and the
longer-lived Balmer absorption that we observe. We will discuss
the the possible link between these two populations of E+S0s in more detail
in \S~5.

We additionally searched for environmental trends in the metal line
indices Mg$_2$, Mg$^b$, and [MgFe]$^\prime$, but detect no significant
trends. In contrast with the correlation between, e.g.,
[MgFe]$^\prime$ and $\sigma$, it seems that the overall metallicity of
early-type galaxies is not dependent on environment, but mainly on the
galaxy's formation history, as traced by $\sigma$. While the highest
values of [MgFe]$^\prime$ that we measure tend to be seen in galaxies
within the Virial radius, this is simply a reflection of the fact that
most of the massive cluster galaxies with highest velocity dispersion
are also located in the cluster core.

\subsubsection{Residual Correlations}

\begin{figure}[t]
\centering
\includegraphics[width=\columnwidth]{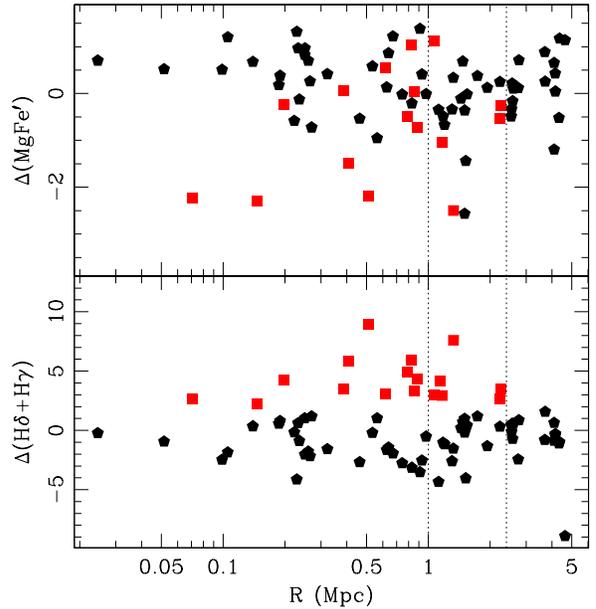}
\caption{\label{fig:balmer_sigma_diff} Residuals from the
  [MgFe]$^\prime$--$\sigma$ and (H$\delta$+H$\gamma$)--$\sigma$
  relations. The interaction with the environment has
  spurred increased star formation within the Virial radius. Galaxies
  that lie above $\sim2$ in the bottom panel have
  Balmer absorption that is unexpectedly strong for their velocity
  dispersion. Similarly, galaxies with $\Delta$[MgFe]$^\prime$ below -1
  are showing unexpectedly weak metal lines. In both panels, galaxies
  with $\Delta($H$\delta$+H$\gamma)>2.0$ are marked with red squares.}
\end{figure}

In addition to the environmental trends in spectral line strengths
discussed above, we can probe environmental effects on the early type
population even more sensitively by examining the residuals from the
[MgFe]$^\prime$--$\sigma$ and Balmer--$\sigma$ relations presented in \S~4.1.
In Figure~\ref{fig:balmer_sigma_diff}, we plot $\Delta$[MgFe]$^\prime$
and $\Delta$(H$\delta$+H$\gamma$) versus projected radius. Similar to
what we saw in Figure~\ref{fig:ew}, we see a population of galaxies
with enhanced Balmer absorption within the Virial radius at 1.7 Mpc.
In this plot, though, we are able to
remove the effects of galaxy mass by looking, in effect, at
galaxies that have excess Balmer absorption {\it compared to other
galaxies at the same velocity dispersion}. This is clear evidence,
then, that the observed Balmer excess is an environmental effect,
and not simply due to a dual correlation between
line strength--mass and mass--cluster radius.

In the top panel of
Figure~\ref{fig:balmer_sigma_diff}, where we plot
$\Delta$[MgFe]$^\prime$, galaxies that also exhibit
$\Delta$(H$\delta$+H$\gamma$) $> 2 \mbox{\AA}$ are marked with red squares.
Virtually all of the galaxies with anomalously
low [MgFe]$^\prime$, which we made note of in
Figure~\ref{fig:mgfe_sig}, belong to this population of
Balmer-strong galaxies. For the first 1~Gyr after a burst
of star formation, we would in fact expect that the A-stars causing
the enhanced Balmer absorption would also tend to fill the metal
absorption lines with emission, as these lines are much weaker in
A-stars than they are in the older stars making up the bulk of the
stellar population. In fact, \citet{tantalo04} suggests that a young
stellar population has just this effect: $[MgFe]^\prime$ is very low
for mean ages less than 2 Gyrs, while $H\beta$, and presumably the
other Balmer indices are enhanced.

We find no evidence that any of Mg$_2$, Mg$^b$, [OII], or
any Balmer lines are correlated with the residuals from the FP,
$\Delta Log(M/L_V)$. Nor is there evidence that the residuals from
any of the Mg--$\sigma$, or Balmer--$\sigma$ relations are correlated
with $\Delta Log(M/L_V)$. This indicates that recent
perturbations to the stellar population of an early-type are somewhat decoupled from its
overall mass to light ratio. While the FP, [MgFe]$^\prime$--$\sigma$,
and Balmer--$\sigma$ relations establish a clear link between the
formation histories of E+S0s and their current appearance, the lack of
correlation between residuals indicates that environmental effects
play a relatively minor role in determining $M/L_V$. The initial time
of formation and mass-assembly history seem to be the dominant factors.

\subsection{Luminosity Trends}
Figure~\ref{fig:OIIfrac} reveals that there is a luminosity trend
in the fraction of E+S0 galaxies with EW([OII]) $<-5$, as well as
the fraction with enhanced Balmer absorption. For the brightest
galaxies, $I < 20.3$, the fraction of [OII]-emitters decreases
monotonically, from the field measurement of \citet{tt02} 
to the cluster core. However, when we
include slightly dimmer galaxies, to $I=21.1$, we see an enhanced
fraction of emitters in the cluster periphery, though [OII]
emission is still rare within $2.4$ Mpc. Now including our full
sample to $I=22.5$, the fraction of star-forming galaxies in both
the transition region and periphery are higher than expected. In
the transition region, the jump represents the addition of the dim
but strongly emitting galaxies discussed above (which have a mean
magnitude of $I=21.1$). The overall trend
suggests that the star formation observed in the outskirts of this
cluster is suppressed earliest in the largest and brightest E+S0s.

A similar trend is seen in the fraction of early types with
(H$\delta$+H$\gamma$) $> -2 \mbox{\AA}$. As discussed above, we see a
jump at all luminosities in the fraction of Balmer-strong galaxies
within the Virial radius, possibly due to the same mechanism that
causes the enhanced [OII] emission. But within each radial zone, the
fraction of galaxies with prominent Balmer absorption increases with
increasing magnitude;the mean magnitude of all such Balmer-strong
early types is $I=20.3$. However, unlike [OII], which probes current star
formation only, the Balmer lines are also sensitive to the stellar population
age. This gradient in Balmer strength with luminosity could simply
reflect an earlier formation time for the brightest galaxies,
regardless of local environment.

The brightest galaxies in our sample ($I<20.3$) follow the FP and show
few signs of star formation. Similarly, galaxies to $I=21.1$ lie on
the FP (albeit with a high scatter) and exhibit star formation mainly in the periphery.
Perhaps star formation is quenched early during infall of these
galaxies, with mostly passive luminosity evolution from that point. 
Because the
dimmest galaxies seem to undergo the most prominent bursts of star
formation (though AGN activity has yet to be ruled out), it seems
that the physical mechanism responsible affects smaller galaxies more
dramatically than larger galaxies. Of course, the elevated fraction of
galaxies with strong Balmer absorption demonstrates that all early
types, bright or dim, are affected to some degree. When evaluating the
physical mechanisms that lead to the observed environmental evolution,
we must also consider this dependence on luminosity.

\section{DISCUSSION}

\label{sec:disc}

Having presented our results in the previous section, we now
proceed to a more comprehensive discussion of our findings trying
to put together the various pieces of information and reconstruct
the evolutionary process, its timescales and the physical
mechanisms driving it. First, in Section~\ref{ssec:time}, we will
combine the radial trends obtained so far and discuss them
jointly. Then, in Section~\ref{ssec:model} we will present a
simple infall model which indicates that a common physical
mechanism may cause several of the radial trends
seen in various subpopulations of E+S0s. In
Section~\ref{ssec:physics} we discuss the physical mechanisms at
work and conclude that our study of early-type galaxies has revealed
the action of at least two different environmental
processes: starvation and/or galaxy harassment is quenching 
low-level star formation over a long timescale in early-types beyond the
cluster core, while harassment and/or interactions with shocks
in the outer parts of the ICM are producing the sudden bursts of
star formation observed in small E+S0s around the Virial radius.

\subsection{Radial trends and star formation timescales}

\label{ssec:time}

\begin{figure*}
\includegraphics[width=2.0\columnwidth]{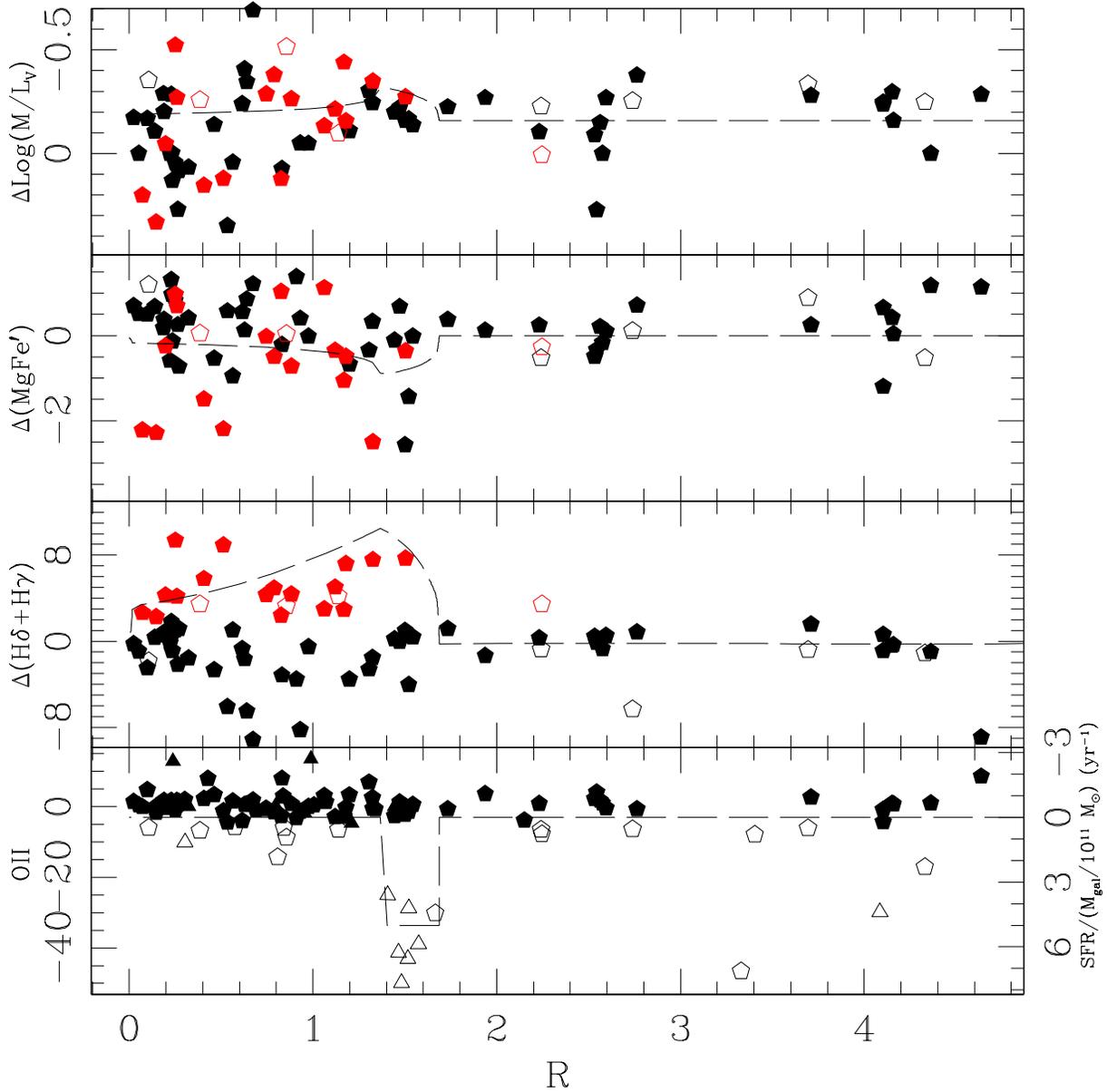}\caption{\label{fig:gal_evol}
  Radial trends in the residuals of dynamical relations. From top to
  bottom, $\Delta Log\left(M/L_V\right)$,
  $\Delta\left(MgFe\prime\right)$,
  $\Delta\left(H\delta+H\gamma\right)$, and $[OII]$ vs. R for
  reference. Galaxies brighter than $I=21.1$ are 
  marked with pentagons, and triangles represent galaxies in the 
  range $21.1<I\le22.5$.
  In the top three panels, we indicate every galaxy exhibiting high
  Balmer absorption ($\Delta\left(H\delta+H\gamma\right) > 2\mbox{\AA}$)
  with a red mark. Likewise, all galaxies with [OII]$<-5\mbox{\AA}$
  in the bottom panel are plotted with open symbols in each panel.
  Overplotted on each plot is the track of a simple
  early-type galaxy model, as the galaxy falls freely from the
  outskirts of the cluster, and has an encounter at the Virial radius
  which triggers a starburst of duration 200Myrs, totaling $1\%$ of
  the galaxy's mass.}
\end{figure*}

In the residuals from the Fundamental Plane, we have observed a
gradient in the mean mass to light ratios of E+S0s, allowing us to
observe the transition from the oldest, highest $M/L_V$ galaxies
in the cluster core, to the galaxies in the cluster periphery with
younger luminosity-weighted ages, nearly matching the values of
$M/L_B$ seen in field E+S0s \citep{tt01,tt05b,vanderwel05}. The scatter in the FP
residuals is unexpectedly high within the inner 1~Mpc of the
cluster, but the FP appears tighter outside of this radius. Direct
measurements of spectral line strengths (Figure~\ref{fig:ew}) and
the fraction of [OII]-emitting early-types
(Figure~\ref{fig:OIIfrac}) both reveal a mildly declining star
formation rate within early-types, from the field toward the cluster
core. This is interrupted by an
interaction at the Virial radius that causes enhanced [OII]
emission in a population of small, dim early types, along with
enhanced Balmer
absorption and depressed [MgFe]$^\prime$ absorption strength for a
population of larger, typical cluster E+S0s.

In Figure~\ref{fig:gal_evol}, we combine the previously shown radial
trends in $\Delta Log(M/L_V)$, $\Delta(MgFe^\prime)$,
$\Delta$(H$\gamma+$H$\delta$), and EW([OII]) into a single plot. As in
Figure~\ref{fig:balmer_sigma_diff}, in the top three panels we
indicate every galaxy exhibiting high Balmer absorption with a red
mark. Likewise, all galaxies with significant [OII] emission in the
bottom panel are plotted with open symbols in each panel. While the
galaxies with unexpectedly low [MgFe]$^\prime$ correlate with the
Balmer-strong galaxies, as mentioned in \S~4 above, there is no
similar correlation with $\Delta Log(M/L_V)$. We see some
Balmer-strong galaxies at both low and high $M/L_V$, which indicates
that the star formation causing the enhanced Balmer line strengths
does not drive the variations in $M/L_V$ that we observe; the bursts
of star formation are perhaps too minor to affect the overall $M/L_V$
of the galaxy, at least to levels detectable in our data.

Briefly, we recap the characteristic timescales for changes in each of the
quantities plotted. 
[OII] emission traces the instantaneous rate of star formation, so
that the observed strong [OII] emitters mark the location of a
``hot spot" of current star formation at 1.7~Mpc radius (though we
emphasize that these objects are not spatially clustered).
Enhanced Balmer absorption will persist for about 1~Gyr after a
burst of star formation, and so there is a time lag of $< 1$~Gyr
between the termination of star formation in these galaxies and
the time when we observe them. Changes in $M/L_V$ should persist
for a similar amount of time after a recent burst of star
formation (until the luminous A stars begin to disappear), though
we expect that $M/L_V$ is less sensitive to low-level star
formation because it depends more closely on the integrated
stellar light of the galaxy, which changes by a smaller amount
than the strong spectral lines.

\subsection{A simple infall model}

\label{ssec:model}

We have so far presented evidence for a seemingly abrupt encounter
with the environment
that triggers star formation in early-types at or near the Virial
radius, along with
a moderate decline in star formation from the periphery to the core,
and a trend in $M/L_V$ that suggests that the stellar populations of
early types at larger radius are younger. 
In this section we present a simple infall model as a tool to 
compare the timescales and strengths of several of the observed
features in the early-type population. Specifically, we wish to
examine the possibility that both the galaxies with strong [OII]
emission and those with excess Balmer
absorption could be caused by a common physical process. And while we
have shown in \S~5.1 that the FP residuals, $\Delta
Log(M/L_V)$, do not correlate with the Balmer-strong galaxies, we will
also use this simple infall model to constrain the level of variation
in $M/L_V$ that could be induced by the environment.

In this
idealized model, we follow an early-type galaxy as it proceeds
toward the cluster center evolving passively, until it reaches
the Virial radius where it undergoes a short burst of star
formation (adding 1\% to the total stellar mass, 200Myrs of
duration). After the burst, the
galaxy continues its infall evolving passively. For this model we
use the prescription of Paper I to convert from infall time to
cluster projected radius: assuming that the galaxy is on its first
infall toward the cluster, it begins with a small velocity at
$R=5$~Mpc and accelerates freely toward the cluster potential.

The observable properties of this model are then computed using
stellar population synthesis models by \citet{bc03} to compare
with observations.  Although this is clearly a simplified model --
for example it neglects ``backsplash'' \citep{gill05, ellingson04}, i.e.
the fact that especially in the cluster center, and even out to two
virial radii, some galaxies might not be on their first approach 
toward the cluster (see discussion
in Paper I), and it is a closed boxed model, i.e. there is no
provision for galaxies being transformed into E+S0s during the
infall -- we will show that this model reproduces  the
strengths and timescales of many of the observed features 
and provides useful guidance to interpret the
observations.

The predictions of the model are shown as a dashed line in
Figure~\ref{fig:gal_evol}. Specifically, we plot the difference in
$\Delta Log(M/L_V)$, $\Delta(MgFe^\prime)$,
$\Delta$(H$\gamma+$H$\delta$), and star formation rate (SFR)
between the infalling model galaxy and an identical model without
the starburst.  Since we are interested only in the predicted change
in line strengths due to a recent star formation event, 
the proper quantity to plot is this difference between the index 
strengths predicted by the passive model and the one with a small 
starburst overlayed.
We emphasize that the model track for $\Delta
Log(M/L_V)$ is meant to replicate only the change in $M/L_V$
caused by a small starburst, and does not account for the radial
gradient in $\Delta<Log(M/L_V)>$ seen in the data.

To calibrate the star formation rates predicted by the models to
the [OII] equivalent widths observed, we used SFRs derived by
\citet[][and private communication]{kodama04} from their deep
H$\alpha$ images, according to the procedure outlined in \S~3.4.

The choice of a 1\% starburst produces a SFR and [OII] equivalent
width equal to that observed in the largest/brightest of our 
low-luminosity [OII] emitting galaxies, which we estimate to have
a stellar mass up to around $5 \times 10^{10}$ M$_\odot$ (See \S~3.4). 
The 200~Myr timescale approximately matches the spread in cluster 
radius over which we see these [OII] emitters. The
stellar mass associated with the starburst could vary with galaxy
size and need not add up to 1\% of the galaxy mass. For our [OII] emitters, 
the luminosities and equivalent widths imply that up to a 5$\%$ burst 
could be involved, depending on its duration. However, 1\% represents
a good upper limit for the larger galaxies with enhanced Balmer 
absorption, as we would otherwise expect to observe ``post-starburst'' 
k+a spectra; the excess Balmer absorption is not strong enough to 
place any in this category.

Examining the track of the model in Figure~\ref{fig:gal_evol}, it
is remarkable how well the expected change in Balmer line strength
matches the observations in both maximum strength and in the time
it takes this enhanced absorption to decay away. This lends
strength to our contention that the enhanced Balmer absorption and
the high [OII] emission are reflections of the same physical
process.

We also observe a dip in the strength of [MgFe]$^\prime$ just where we
would expect it, though the magnitude of this dip does not seem great
enough to account for the observed [MgFe]$^\prime$ decrement. This may
simply be a deficiency in the \citet{bc03} models, as the 
observed values of [MgFe]$^\prime$ span a larger range than could be
predicted by the models at any SSP age or metallicity. 
In contrast, the models easily account 
for the full range of observed Balmer line strengths. Therefore, changes
in [MgFe]$^\prime$ due to a small starburst could be under predicted by
the \citet{bc03} models.

We note that the population of [OII] emitters observed cannot
  evolve directly into the population of galaxies with strong Balmer lines,
though they could still be indicators of the same
physical mechanism. 
The Balmer-strong E+S0s are brighter and larger in effective radius,
  and any fading of the [OII] emitters would only increase the 
difference between these populations. Since they are near our 
spectroscopic magnitude
limit, we would not expect to observe faded remnants of the
[OII] emitting galaxies. On the other hand, the lack of bright early
types with strong [OII] emission might arise because the 
timescale for the Balmer lines to decay (1~Gyr) exceeds the starburst 
timescale (200~Myrs), so that we would not
  expect to observe more than one of the emission line progenitors of these
  Balmer-strong galaxies. 
  This is partly due to the overall
  smaller number of bright E+S0s observed within the transition region
  (1--2.4~Mpc), in comparison to the $\sim50$ observed in the cluster
  core.

\begin{figure*}[t]
\centering
\includegraphics[height=4in]{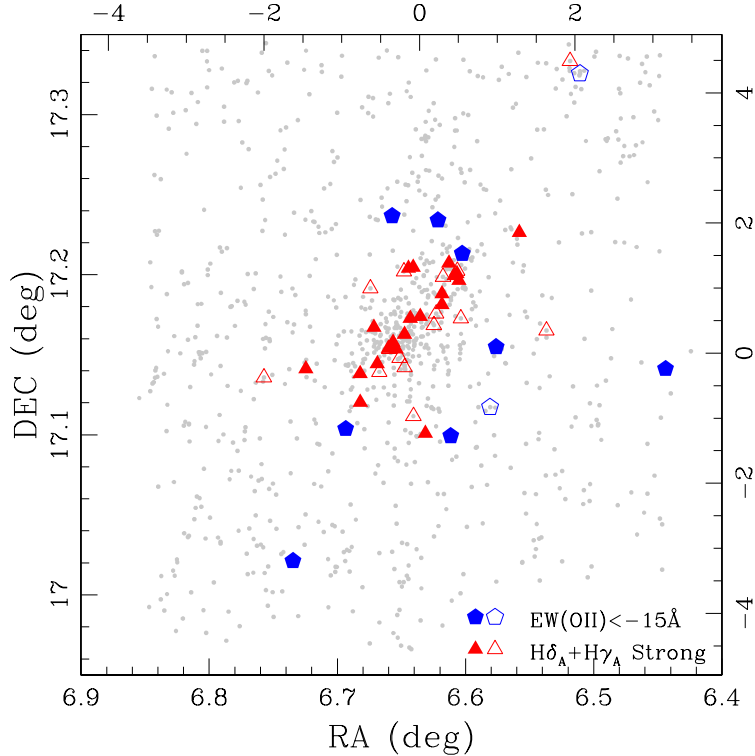}
\caption{\label{fig:spatial_dist} Spatial distribution of [OII]
  emitters and galaxies with excess Balmer absorption, in comparison
  to the overall distribution of cluster members in the combined
  spectroscopic and photometric (Smith et al. 2005) redshift catalog
  (gray dots). Blue pentagons are strong [OII] emitters, with two open
  symbols marking the merger system and possible spiral interloper
  seen in Figure~9. Filled red triangles mark galaxies where the
  Balmer--$\sigma$ residuals are 
  $\Delta\left(H\delta+H\gamma\right) > 2\mbox{\AA}$, and unfilled
  triangles represent galaxies without measured velocity dispersions, with 
  EW(H$\delta_A+$H$\gamma_A) > -2\mbox{\AA}$. Top and right axes show
  projected radius from the cluster center, in Mpc.}
\end{figure*}

To further examine our hypothesis that the [OII]-emitting 
and Balmer-strong galaxies are caused by the same environmental 
interaction during infall, we can examine the 2D spatial distribution 
of these galaxies. In Figure~\ref{fig:spatial_dist}, we plot [OII] 
emitters in blue and early-types with strong Balmer absorption in red, 
on top of the overall distribution of cluster members from the combined
spectroscopic and photometric redshift catalogs. The [OII]
emitters and Balmer absorbers follow the overall distribution
of cluster members closely, lying mostly along the major
axis of the cluster (stretching from the overdensity NW of the
cluster core through to the SE side of the core). Though the [OII] 
galaxies perhaps extend in a larger arc at $\sim1.7$~Mpc, their small 
number makes it difficult to determine their azimuthal
distribution. The distribution supports the idea
that they have been perturbed during infall, as this is most likely 
occurring predominantly along the NW to SE path.

Turning to the top panel in Figure~\ref{fig:gal_evol}, it seems that
the small starbursts at the Virial radius cannot alter the overall
$M/L_V$ of early types by a large enough amount to account for the
large scatter that we see in the cluster core.
It seems that there must simply exist a mix of
older, high $M/L_V$ galaxies and more recently arrived low $M/L_V$
galaxies in this region. On the other hand, the triggered star
formation may help prolong the length of
time that an infalling E+S0 remains at low $M/L_V$ after reaching the
cluster core. This could possibly explain why the scatter in $M/L_V$
seen in this cluster is so much higher than that observed in Cl~1358+62 at
$z=0.33$ \citep{kelson00c}; the mechanism causing this star formation
in Cl~0024 might not be active or significant in all clusters at this redshift.

While the simple model presented here does not attempt to account
for the observed radial gradient in $M/L_V$, we note that the
difference in $Log(M/L_V)$ between a galaxy at 5~Mpc radius, of
age 2.7~Gyr, and that same galaxy 5~Gyr later (corresponding to
the total free-fall time to the cluster center), is equal to
-0.35. This is nearly identical to the observed difference in
$Log(M/L_V)$ between the mean value in the periphery, and the
$M/L_V$ of the oldest galaxies in the core. While the galaxies in
the periphery are almost certainly not this young, the observed
gradient in $M/L_V$ seems consistent with a difference in
formation age (or in time since the last major burst, c.f. Bower
et al. 1998 and Treu et al. 2005b) of a few Gyrs between cluster
periphery and core.  A gradient in formation age, then, is a
viable explanation for the $M/L_V$ gradient we observe, though the
gradual decline in star formation that we observe could also be
important in generating the $M/L_V$ gradient.

Having suggested that the observed spectral features of our early-type
  sample are due to a small starburst, it is natural to ask if
  such a burst is realistic; in other words, do E+S0s at redshift
  $\sim0.4$ contain enough residual gas to undergo such episodes of
  star formation? 
  According to \citet{bettoni03}, the 
  fraction of molecular gas
  detected in local ``normal'' early type galaxies (i.e. showing no
  signs of current or recent interaction or dwarf cannibalism, but
  otherwise selected across all local environments), can
  approach 1\% of the total stellar mass (with typical values ranging
  from 0.05--0.5\%). At intermediate redshifts, some field
  E+S0s have peculiar color gradients that may indicate recent star 
  formation \citep[e.g.][]{menanteau04}. Allowing for the possibility 
  of faint, undetected
  interaction features in our much more distant sample, as well as the
  expected decrease in the remnant molecular gas in ellipticals since
  $z=0.4$, we expect that a small burst of star formation of up to 1\%
  is entirely plausible for early-type galaxies in Cl~0024. 
  Additional evidence comes from the observation 
  \citep[e.g.][]{tt05b, bressan96} that the combinations of 
  spectral line strengths observed in early-type galaxies are best
  explained if most early-types have undergone secondary episodes of
  star formation at some point in their past; the Balmer-strong
  galaxies observed here could possibly have undergone such secondary 
  bursts in the last 1~Gyr.

\subsection{Physical Mechanisms}

\label{ssec:physics}

The early-type galaxies in Cl~0024 have proven to be effective
signposts for identifying two possible forms of environmental
evolution in this cluster, but can we identify the
physical processes responsible?  We observe a slow
decline in remnant star formation from the field/periphery to the core
which may indicate a physical mechanism at work which slowly
quenches star formation. Seemingly overlaid on top of this mild
trend is a nearly-instantaneous interaction that triggers a small
burst of star formation in many of these early-type galaxies at
the virial radius.

First, we address physical mechanisms that could be responsible
for the gradual decline in star formation rate toward the cluster
core, best illustrated in Figure~\ref{fig:OIIfrac}. In analyzing
the cause of the observed decline, we must also consider the
morphology--density (T--$\Sigma$) relation observed in Cl~0024. In
Paper I, it was argued that the existence of the T--$\Sigma$
relation across several orders of magnitude in projected local
density indicates that whatever process causes the morphological
segregation of galaxies by environment, it acts slowly. The slow
``starvation''(See Paper I) of galaxies during infall is
consistent with the existence of the T--$\Sigma$ relation, as is the
action of galaxy harassment \citep{moore96, moore99, mlk98}.

Starvation encompasses several physical mechanisms that all serve
to slowly deprive a galaxy of cold gas available for star
formation. For example, thermal evaporation of the galaxy
interstellar medium (ISM) by interaction with the hot cluster ICM
\citep{cowie77}, or turbulent and viscous stripping of the ISM
\citep{nulson82, tonniazo01} are two possible causes of
starvation. Either of these processes could be acting in Cl~0024,
and are consistent with both the observed gradient in star
formation rate and the existence of the T--$\Sigma$
relation. 

\citet{kodama04}, however, argue that the truncation of star
formation in cluster galaxies must be rapid, based on their
observation that the H$\alpha$ luminosity function does not vary with
environment. Both our paper and theirs find a similar decline in the
{\it fraction} of star forming galaxies, but they do not observe a
decline in the mean strength of emission. The discrepancy may
be due to the different levels of star formation probed: the 
low levels of remnant star formation in the early-type population 
discussed here may be quenched slowly, while a separate physical 
process (such as harassment) could
rapidly decrease the moderate star formation observed 
by \citet{kodama04}, which occurs mostly in the spiral population. 
Combined with a longer timescale for morphological
transformation, this could lead to the ``passive spirals'' they
discuss. 

Paper I also argued that the existence of the T--$\Sigma$ relation
means that the morphological mix is set by the pre-cluster
environment: the overall formation history within assembling
groups is the dominant factor in determining the morphological mix
seen in the cluster. This view is consistent with the trend we
observe in the $M/L_V$ ratios of cluster early-types. We find that
the differences in $M/L_V$ must arise partially out of variations
in the mass--assembly history of early-type galaxies, though the 
radial trend in $<\log(M/L_V)>$ may also reflect the quenching of
remnant star formation during infall.

Interrupting the slow decline in star formation are the small
bursts of star formation we observe. Rapidly acting mechanisms
that could trigger such a burst include galaxy harassment or some
encounter with the ICM. As noted in \S~4,
major mergers are ruled out because the galaxies seem to be in low
density regions, and all but one of the galaxies involved show no visual
evidence for such a disturbance.

Galaxy harassment becomes important near the Virial radius as both
galaxy crowding and the typical velocity of an interaction become
high enough. (See Paper I.) Further, it is a rapid interaction,
and the galaxy that triggers one of our observed starbursts can
quickly move away from the vicinity of the starburst galaxy.
\citet{moore99} present simulations of the harassment of
small Sc/Sd spiral galaxies as they fall into the cluster. Such
galaxies can be quickly transformed into objects that appear to have an
early-type morphology. These harassment remnants are small objects
($R_e=1-3$kpc), similar to our observed [OII] emitters, and could 
represent the origin of the large population of dwarf spheroidals 
observed in clusters today. 
In addition, \citet{moore99} predict that harassment would drive 
the gas within a spiral toward the galaxy's center, thus providing 
a natural mechanism for fueling the starbursts
(or AGN) we observe. 

If these are the remnants of harassed spirals, they may
not be on their first infall into the cluster; the timescale for
passage from the cluster core back out to the virial radius may be short
enough ($<1$~Gyr) that star formation induced by interactions with large
ellipticals in the cluster center could still be visible . A possible
explanation, then, for finding them preferentially at the virial
radius could be that this is near the apse of
their orbit within the cluster. 

Under the harassment scenario, we speculate that 
the population of Balmer-strong 
galaxies could represent the large cluster 
ellipticals which cause the 
harassment, perhaps cannibalizing young stars from smaller 
harassed galaxies. Alternatively, they could be undergoing 
harassment themselves, albeit with a much smaller gas reservoir 
and a correspondingly smaller episode of star formation. 

Ram-pressure stripping or triggering of star formation has been
thought to be effective only much closer to the cluster core, and
should not be significant near the Virial radius where the
environmental interaction is taking place. However, recent work on
large-scale shocks and particularly the Virial shock--the shock
boundary between the hot, dense ICM and the colder, diffuse
infalling gas from outside the cluster--suggests that a galaxy
passing through this shock boundary could be given enough of an
impulse to trigger stripping of HI gas and the collapse of any
molecular clouds \citep{evrard91, quilis00, ryu03, furl04}.
Particularly relevant to Cl~0024 is the work by \citet{roet96}.
They showed that a merger between two clusters or sub-cluster
clumps can cause powerful shocks to propagate through the ICM. As
the redshift distribution of galaxies in Cl~0024 is double-peaked
\citep{czoske}, indicating a large infalling group or sub-cluster,
such strong shocks could be responsible for the triggered
starbursts we observe. If this is the case, we would not expect to
see such bursts in other clusters with more regular structure.

Our sample of 104 E+S0s contains only four galaxies from the
smaller Peak B of the redshift distribution in Cl~0024
(Figure~\ref{fig:redshifts}), so we are unable to more carefully
explore possible differences between early type galaxies in these
two peaks. But in the local universe, \citet{poggianti04} finds a
population of star forming dwarf galaxies that lie along contours
in the X-ray substructure of the Coma cluster. It is possible that
the [OII] emitters in Cl~0024 lie along similar boundaries in the
cluster substructure, but the nearly face-on orientation of the
merger between Peak A and Peak B makes testing this hypothesis
difficult. In this scenario, the existence of Balmer-strong
galaxies interior to the Virial radius would imply that some sort
of boundary shock between Peak A and Peak B has persisted for at
least $\sim1$~Gyr, the timescale for the excess Balmer absorption
to decay away. 

Our small population
  of strong [OII] emitters is a signpost to an important physical
  interaction, and is not a peculiarity unique to this cluster. 
  A similar population of dim, compact, E+S0s with [OII] emission 
  has also been reported by \citet{homeier}, for a galaxy cluster 
  at $z=0.84$. While the cluster they study is not virialized, the
  [OII] emitters they observe seem to reside in a region outside
  of the contours of the cluster X-ray emission. And work by
  \citet{ebeling} examining the radial distribution of X-ray point 
  sources in the MACS sample of clusters has uncovered a prominent
  spike in the number of cluster AGN detected in the area around the
  Virial radius ($\sim2.5$~Mpc in the clusters studied). These AGN may
  be stimulated by the cluster environment via the same physical 
  mechanism causing the [OII]-strong early-types. Therefore, even
  though they only represent a small fraction of the total mass and
  star-formation in clusters, these galaxies warrant further study of
  their role as indicators of environmental interaction.

While the exact physical mechanisms responsible for the environmental
evolution are still not known, further work on this cluster, and a
similar planned study of the X-ray bright cluster MS~0451-03 at
$z=0.54$ may allow us to determine which processes are
dominant. Analyzing the star formation rates and rotation curves of
spirals in Cl~0024 may help us trace the gradual decline in star
formation rate hinted at here, as star formation in spirals is in
general much more vigorous than in the early types studied so
far. Other authors have found similar declines in the star formation
rate in spiral populations of clusters at intermediate redshift
\citep{abraham96, balogh99, balogh00}. Rotation curves have
also been used to study variations in the kinematics and mass to light 
ratios of spirals, in distant clusters by, e.g., 
\citet{bamford05, ziegler03, milvangjensen03}, and in
Cl~0024 by \citet{metevierkoo}.

Likewise, the X-ray luminosity of MS~0451 is much higher than that
of Cl~0024. If shocks in the ICM really do cause the bursts of star
formation observed in early types near the Virial radius, then we
should see a similar or even stronger effect in MS~0451, but at a
higher radius. If the
strength of the ICM shocks in Cl~0024 is enhanced by the
sub-cluster merger, however, we might expect to see no
starbursting early types at all in MS~0451. The importance of
galaxy harassment should be nearly the same between the two
clusters, so we would expect to see similarly strong starbursts in
MS~0451 if the Cl~0024 bursts are due to harassment.

\section{Summary}

In this paper we have presented the results of an extensive
spectroscopic survey of the cluster Cl~0024+1654, undertaken with
DEIMOS on {\it Keck II} and LRIS on {\it Keck I}. We examine the
detailed spectral and photometric properties of cluster early types
across an area 10 Mpc in diameter. The principal goal of this wide
area survey is to examine variations in galaxy properties with local
environment, in order to identify the physical processes that may
affect the star formation properties or morphological characteristics
of infalling galaxies. {In this paper, we have used the early-type
galaxy population as sensitive indicators of interaction with 
the cluster environment.} Our main observational results are: \\

1) Combining our new redshift survey of nearly 1000 galaxies with
   previous work on this cluster, we present a comprehensive redshift
   catalog for Cl~0024, consisting of 1394 unique objects with
   redshifts, of which 486 are cluster members. From our survey, we
   select a sample of 104 E+S0 cluster members brighter than $I=22.5$,
   for which we can measure spectral line indices. For a subsample of
   71 of these galaxies, with high quality spectra available and $I\le
   21.1$, we also measure reliable velocity dispersions and analyze
   surface photometry from their {\it HST} images.

2) By constructing the Fundamental Plane of Cl~0024, we observe that
   E+S0s at $z\sim0.4$ still have old stellar populations; from a
   comparison with the FP of the Coma cluster ($z\sim0.02$), we infer
   an evolution in mass to light ratio of $\Delta(Log (M/L_V)) = -0.14
   \pm 0.02$.  While the mean change in $M/L_V$ is in line with that
   expected by passive evolution of an old stellar population, the
   high scatter of $40\%$ in $M/L_V$ suggests a more complex assembly
   history for this cluster. The high scatter may be an effect of the
   cluster merger currently underway \citep{czoske}, though at
   least one other intermediate redshift cluster \citep{wuyts04}
   exhibits a similarly high scatter.

3) We observe radial trends in the mass to light ratios of individual
   early types, with the oldest galaxies located in the cluster core
   ($<\Delta Log (M/L_V)>_{R<1Mpc}=-0.13\pm0.04$). Galaxies seem to be
   younger at higher radius, with E+S0s in the cluster periphery
   ($<\Delta Log (M/L_V)>_{R>2.4Mpc}=-0.24\pm0.02$) nearly matching
   the values of $M/L_V$ seen in the field at this redshift
   \citep{tt05b}. We therefore have bridged the gap between the
   observed properties of cluster and field early types at
   intermediate redshift. Similar trends are seen as a function of
   local density. Some fraction of this radial gradient could be caused by the
   ``downsizing" effect described by, e.g.,
   \citet{tt05b,holden05,vanderwel05}. 
   But even when we select a subsample of galaxies within a
   narrow range of galaxy mass ($10.9 < Log(M/M_\odot) \le 11.4$), we
   observe a significant difference in age (or time since the latest burst)
   between galaxies in the cluster periphery and core,
   equivalent to $\delta<Log (M/L_V)>=0.13\pm0.07$. Physically, this trend
   could be explained by some combination of a gradient in formation
   age and variations in recent star formation, perhaps modulated
   by ``starvation".

4) Around the virial radius, we observe a number of small galaxies
   undergoing a burst of star formation, indicated by strong [OII]
   emission. Further, we observe enhanced Balmer absorption for many
   galaxies interior to this radius. This is consistent with an infall
   scenario where, after a starburst, Balmer absorption from young
   stars decays away over the course of $\sim1$~Gyr, as the galaxy
   continues toward the cluster core. The radial distribution of
   enhanced-Balmer E+S0s is consistent with the expected infall
   timescale of such galaxies. Galaxy mergers are not the source of this
   enhanced activity; the [OII] emitters reside in relatively low density
   regions, and do not appear to reside in groups. The small starbursts are
   likely caused by a rapidly acting physical process: possibilities
   include galaxy harassment and shocks in the ICM, perhaps generated by the
   sub-cluster merger in Cl~0024.

5) While the overall early type population is older in the cluster
   core than in the outskirts, we see trends with luminosity that
   indicate that smaller early types have more active star formation
   than larger ones. This seems to confirm evidence for downsizing
   seen by \citet{tt05}, and indicates that galaxy mass is a primary
   indicator of star formation activity, even as environmental effects
   work on top of this to produce radial variations in star
   formation. Both of these ultimately serve to decrease star
   formation toward the highest density regions, though the precise
   link to morphological transformation is still unclear because of the
   different timescales involved.  \\

We have uncovered environmental processes working on cluster early
types, but we have also seen variations in galaxy properties across
the cluster that indicate that both galaxy mass and environment of
initial formation are important in determining the overall assembly
history of early type galaxies.

By obtaining high quality spectra of cluster members across a large
area around Cl~0024, and combining with a panoramic {\it HST} mosaic,
we have been able to uncover physical mechanisms at work that never
would have been seen if we only looked at the cluster core. This
underscores the importance of looking wide as well as far when
attempting to understand how galaxies evolve within the large scale
structure of the universe.

\begin{acknowledgements}
 
We thank Michael Cooper and Jeffery Newman for help with the
DEIMOS data reduction pipeline, F. Owen, A. Metevier, and D.
Koo for sharing redshifts, T. Kodama for allowing us access to 
H$\alpha$ equivalent widths and star formation rates from their
survey, and J. Richard for assisting with DEIMOS observations in
2003. We thank M. Balogh, E. Ellingson, B. Holden, B. Poggianti, A. Babul 
and A. van der Wel for valuable discussions. Faint object
spectroscopy at Keck Observatory is made possible with
LRIS and DEIMOS thanks to the dedicated efforts of J. Cohen,
P. Amico, S. Faber and G. Wirth. We acknowledge use 
of the Gauss-Hermite Pixel Fitting Software developed by R. P.
van der Marel. The analysis pipeline used to reduce the
DEIMOS data was developed at UC Berkeley with support from
NSF grant AST-0071048. TT acknowledges financial support from
a Hubble Fellowship grant HST-01.167.01. RSE acknowledges
financial support from NSF grant AST-0307859 and STScI grants 
HST-GO-08559.01-A and HST-GO-09836.01-A.
 
\end{acknowledgements}

\clearpage
\LongTables
\begin{deluxetable*}{cccccccc}
\centering
\tablewidth{0pt}
\tablecaption{Observed E+S0 cluster members}
\tablenum{1}
\tabletypesize{\footnotesize}

\tablehead{\colhead{Object} & \colhead{$\alpha$} & \colhead{$\delta$} &
  \colhead{$z$} & \colhead{F814W} &
  \colhead{R} & \colhead{$\Sigma_{10}$} & \colhead{Morph} \\
\colhead{} & \colhead{($^o$)} & \colhead{($^o$)} & \colhead{} &
\colhead{ (mag) } & \colhead{(Mpc)} & \colhead{(Mpc$^{-2}$)} & \colhead{} }

\startdata
p0i102c3 & 6.65796 & 17.16050 & 0.3897 & 19.42 & 0.14 & 255 & S0 \\
p0i139c3 & 6.66004 & 17.16630 & 0.3974 & 20.15 & 0.19 & 300 & E/S0 \\
p0i15c3 & 6.64000 & 17.15560 & 0.3984 & 19.84 & 0.27 & 195 & S0 \\
p0i170c3 & 6.64412 & 17.17120 & 0.3931 & 19.24 & 0.22 & 214 & E \\
p0i1c3 & 6.65400 & 17.15350 & 0.3940 & 20.23 & 0.20 & 257 & E/S0 \\
p0i1c4 & 6.66014 & 17.15402 & 0.3967 & 20.54 & 0.25 & 212 & E \\
p0i206c3 & 6.64763 & 17.16280 & 0.3972 & 18.51 & 0.07 & 951 & E \\
p0i209c3 & 6.64012 & 17.16750 & 0.3893 & 21.76 & 0.24 & 214 & S0 \\
p0i217c3 & 6.65629 & 17.16270 & 0.3912 & 20.18 & 0.10 & 254 & E/S0 \\
p0i24c2 & 6.65271 & 17.14714 & 0.3954 & 21.11 & 0.32 & 204 & E/S0 \\
p0i38c4 & 6.66325 & 17.15350 & 0.3860 & 20.36 & 0.30 & 244 & S0 \\
p0i39c3 & 6.64025 & 17.15860 & 0.3919 & 19.62 & 0.23 & 265 & S0 \\
p0i3c3 & 6.65633 & 17.15240 & 0.3915 & 19.64 & 0.24 & 255 & S0 \\
p0i42c3 & 6.65792 & 17.15610 & 0.3994 & 18.85 & 0.19 & 390 & E \\
p0i45c2 & 6.65133 & 17.14800 & 0.3951 & 21.46 & 0.30 & 183 & S0 \\
p0i49c2 & 6.64896 & 17.14980 & 0.3946 & 19.26 & 0.27 & 200 & S0 \\
p0i4c4 & 6.65950 & 17.15350 & 0.3910 & 20.95 & 0.25 & 212 & E \\
p0i53c4 & 6.67150 & 17.16720 & 0.3976 & 18.88 & 0.41 & 102 & E \\
p0i55c2 & 6.64754 & 17.14220 & 0.3960 & 19.91 & 0.43 & 206 & S0 \\
p0i66c3 & 6.65321 & 17.15800 & 0.3843 & 20.01 & 0.11 & 372 & S0 \\
p0i72c3 & 6.64875 & 17.16200 & 0.3920 & 18.11 & 0.05 & 1024 & E \\
p0i79c3 & 6.65000 & 17.16280 & 0.3883 & 17.75 & 0.02 & 637 & E \\
p0i85c3 & 6.65663 & 17.15780 & 0.3943 & 19.02 & 0.15 & 339 & E \\
p0i91c3 & 6.64575 & 17.17260 & 0.3887 & 19.57 & 0.23 & 151 & E \\
p0i95c3 & 6.64313 & 17.17280 & 0.3876 & 18.77 & 0.26 & 274 & E \\
p10i1c2 & 6.78233 & 17.18040 & 0.3967 & 18.66 & 2.59 & 14 & S0 \\
p11i147c4 & 6.72475 & 17.17760 & 0.3955 & 20.36 & 1.47 & 16 & S0 \\
p12i160c2 & 6.67425 & 17.19140 & 0.3951 & 20.40 & 0.74 & 44 & E \\
p12i160c3 & 6.64088 & 17.20460 & 0.3955 & 19.79 & 0.88 & 88 & E/S0 \\
p12i168c4 & 6.65725 & 17.23670 & 0.3800 & 21.55 & 1.52 & 20 & E/S0 \\
p12i73c4 & 6.66853 & 17.23410 & 0.3962 & 20.14 & 1.50 & 22 & S0 \\
p13i130c3 & 6.62154 & 17.23410 & 0.3934 & 21.98 & 1.57 & 25 & E/S0 \\
p13i133c4 & 6.59632 & 17.20980 & 0.3967 & 21.42 & 1.45 & 57 & E \\
p13i135c4 & 6.59546 & 17.20870 & 0.3965 & 18.61 & 1.44 & 56 & S0 \\
p13i1c4 & 6.60596 & 17.22300 & 0.3975 & 19.88 & 1.52 & 34 & E \\
p13i25c2 & 6.58954 & 17.23660 & 0.3962 & 19.59 & 1.94 & 17 & S0 \\
p13i78c4 & 6.61300 & 17.20720 & 0.3973 & 18.93 & 1.18 & 109 & E \\
p13i86c4 & 6.60258 & 17.21320 & 0.3929 & 22.27 & 1.41 & 60 & E/S0 \\
p14i2c2 & 6.55800 & 17.22640 & 0.3933 & 20.61 & 2.25 & 13 & E \\
p15i144c4 & 6.47717 & 17.27400 & 0.3967 & 19.99 & 4.10 & 7 & S0 \\
p17i2c4 & 6.75733 & 17.13590 & 0.3941 & 20.92 & 2.15 & 8 & S0 \\
p18i43c2 & 6.72658 & 17.14080 & 0.3925 & 19.01 & 1.54 & 31 & E \\
p18i51c2 & 6.72458 & 17.14120 & 0.3933 & 19.87 & 1.50 & 26 & E \\
p18i66c3 & 6.71600 & 17.14440 & 0.3934 & 19.98 & 1.32 & 26 & E \\
p19i1c3 & 6.58353 & 17.17150 & 0.3939 & 20.48 & 1.34 & 18 & E/S0 \\
p19i75c4 & 6.60362 & 17.17270 & 0.3941 & 21.41 & 0.96 & 38 & E/S0 \\
p19i92c2 & 6.57613 & 17.15490 & 0.3811 & 21.15 & 1.48 & 22 & E/S0 \\
p20i35c4 & 6.53687 & 17.16520 & 0.3971 & 19.42 & 2.24 & 8 & S0 \\
p20i48c3 & 6.54025 & 17.18710 & 0.3921 & 20.12 & 2.23 & 7 & E \\
p24i115c3 & 6.61158 & 17.09930 & 0.4082 & 21.65 & 1.52 & 19 & S0 \\
p24i1c3 & 6.63138 & 17.10100 & 0.3981 & 18.87 & 1.33 & 26 & E/S0 \\
p24i42c4 & 6.63029 & 17.11800 & 0.4018 & 20.64 & 1.01 & 42 & E \\
p24i79c4 & 6.64067 & 17.11160 & 0.3927 & 20.64 & 1.07 & 35 & E \\
p24i87c3 & 6.62296 & 17.10520 & 0.3972 & 19.32 & 1.30 & 26 & S0 \\
p25i29c3 & 6.57688 & 17.11700 & 0.3924 & 20.32 & 1.73 & 14 & E \\
p25i93c3 & 6.58088 & 17.11730 & 0.3970 & 19.21 & 1.66 & 19 & S0 \\
p27i174c3 & 6.46279 & 17.14510 & 0.3978 & 19.06 & 3.71 & 5 & S0 \\
p27i9c2 & 6.44396 & 17.14130 & 0.3792 & 21.89 & 4.09 & 5 & E/S0 \\
p28i87c2 & 6.73458 & 17.02130 & 0.3950 & 20.64 & 3.33 & 9 & E \\
p35i112c2 & 6.51850 & 17.33350 & 0.3833 & 20.12 & 4.36 & 21 & S0 \\
p35i143c2 & 6.49954 & 17.33630 & 0.3951 & 19.94 & 4.64 & 15 & S0 \\
p35i43c3 & 6.52433 & 17.32190 & 0.3927 & 19.13 & 4.10 & 33 & E \\
p35i6c3 & 6.51429 & 17.31730 & 0.3946 & 18.85 & 4.15 & 23 & E \\
p35i78c2 & 6.51071 & 17.32570 & 0.3931 & 19.23 & 4.33 & 21 & S0 \\
p35i82c3 & 6.52896 & 17.32850 & 0.3956 & 19.28 & 4.16 & 23 & S0 \\
p36i106c3 & 6.61754 & 17.19870 & 0.3970 & 22.05 & 0.99 & 72 & S0 \\
p36i146c3 & 6.60642 & 17.20250 & 0.3987 & 19.72 & 1.20 & 118 & S0 \\
p36i182c2 & 6.62129 & 17.17350 & 0.3959 & 19.28 & 0.63 & 169 & S0 \\
p36i182c3 & 6.60512 & 17.19650 & 0.3930 & 20.65 & 1.14 & 106 & S0 \\
p36i184c2 & 6.61846 & 17.18120 & 0.3955 & 19.19 & 0.75 & 114 & S0 \\
p36i187c3 & 6.60879 & 17.19940 & 0.3965 & 20.17 & 1.12 & 126 & E/S0 \\
p36i1c4 & 6.63014 & 17.20229 & 0.3907 & 19.45 & 0.91 & 77 & E/S0 \\
p36i2c2 & 6.63088 & 17.18210 & 0.3958 & 18.01 & 0.56 & 99 & E \\
p36i2c3 & 6.61833 & 17.18800 & 0.3922 & 19.92 & 0.83 & 130 & S0 \\
p36i31c4 & 6.63221 & 17.20670 & 0.3905 & 18.73 & 0.97 & 123 & E \\
p36i32c4 & 6.63254 & 17.19910 & 0.3916 & 20.14 & 0.83 & 117 & S0 \\
p36i34c2 & 6.63088 & 17.17420 & 0.3950 & 20.91 & 0.46 & 124 & S0 \\
p36i36c4 & 6.63254 & 17.19970 & 0.3949 & 20.96 & 0.84 & 121 & E \\
p36i3c2 & 6.62933 & 17.18590 & 0.3931 & 20.22 & 0.64 & 105 & E \\
p36i48c3 & 6.61658 & 17.19380 & 0.3934 & 19.74 & 0.93 & 103 & S0 \\
p36i5c2 & 6.63542 & 17.17390 & 0.3969 & 19.18 & 0.39 & 186 & S0 \\
p36i67c4 & 6.64803 & 17.20200 & 0.3951 & 20.87 & 0.81 & 65 & E/S0 \\
p36i70c4 & 6.64450 & 17.20400 & 0.3975 & 20.38 & 0.86 & 75 & E \\
p36i71c3 & 6.60769 & 17.20461 & 0.3996 & 21.25 & 1.21 & 105 & E/S0 \\
p36i76c2 & 6.62469 & 17.16852 & 0.3959 & 20.29 & 0.53 & 151 & S0 \\
p36i76c3 & 6.60667 & 17.20060 & 0.3932 & 19.95 & 1.17 & 124 & S0 \\
p36i89c2 & 6.62292 & 17.17570 & 0.3977 & 19.97 & 0.62 & 154 & S0 \\
p37i110c2 & 6.68208 & 17.12020 & 0.3954 & 20.86 & 1.06 & 30 & S0 \\
p37i122c2 & 6.69346 & 17.10380 & 0.3918 & 21.69 & 1.47 & 16 & E/S0 \\
p37i16c4 & 6.68229 & 17.13830 & 0.3975 & 18.02 & 0.79 & 81 & S0 \\
p37i206c4 & 6.68004 & 17.14495 & 0.3965 & 20.44 & 0.67 & 87 & E \\
p37i207c4 & 6.68607 & 17.14283 & 0.3953 & 21.47 & 0.80 & 146 & S0 \\
p37i64c4 & 6.66879 & 17.14430 & 0.3977 & 20.18 & 0.51 & 72 & E/S0 \\
p37i9c4 & 6.66721 & 17.13930 & 0.3999 & 20.81 & 0.57 & 60 & S0 \\
p38i1c4 & 6.62454 & 17.14090 & 0.4003 & 20.39 & 0.69 & 55 & E \\
p38i76c4 & 6.62533 & 17.13030 & 0.3976 & 20.88 & 0.84 & 37 & S0 \\
p3i72c2 & 6.63067 & 17.32760 & 0.4007 & 20.67 & 3.40 & 7 & S0 \\
p5i9c3 & 6.74192 & 17.26600 & 0.3904 & 20.69 & 2.76 & 10 & E/S0 \\
p7i128c3 & 6.63150 & 17.28690 & 0.3934 & 19.85 & 2.57 & 72 & S0 \\
p7i133c3 & 6.63194 & 17.28540 & 0.3967 & 20.82 & 2.54 & 52 & S0 \\
p7i136c3 & 6.63197 & 17.28490 & 0.3960 & 19.93 & 2.53 & 52 & S0 \\
p7i1c2 & 6.60829 & 17.28990 & 0.3981 & 18.64 & 2.74 & 24 & E \\
p7i67c3 & 6.62904 & 17.28590 & 0.3961 & 20.60 & 2.56 & 71 & S0 \\
p9i20c3 & 6.83968 & 17.16743 & 0.3953 & 19.01 & 3.69 & 9 & S0 \\
\enddata
\tablecomments{Object names are abreviated from those given in
  Paper~I, which listed objects by WFPC2 pointing, WFPC2 chip, and ID
  number. For example, p0i102c3 refers to object number 102 on chip 3,
  from pointing 0. Right ascention ($\alpha$) and declination ($\delta$)
  are for epoch J2000. F814W magnitudes are corrected for galactic
  extinction, adopting $E(B-V)=0.057$ \citep{schlegel98}. 
  $R$ and $\Sigma_{10}$ denote, respectively, the
  projected radius and local density of each galaxy, calculated
  according to the method described in the text. 
  Typical uncertainty in $\Sigma_{10}$ is $25\%$, equal to the rms
  variation in the local density across different measurement
  methods. Errors in F814W magnitudes are less than 0.05 mag rms; more
  information on photometric measurements is given in Paper I.}
\end{deluxetable*}

\clearpage
\LongTables
\begin{deluxetable*}{ccccccccccc}
\centering
\tablewidth{0pt}
\tablecaption{All Measurements of Observed E+S0 members}
\tabletypesize{\scriptsize}
\tablenum{3}

\tablehead{\colhead{Object} & \colhead{OII} & \colhead{H$\delta_A$} &
  \colhead{H$\gamma_A$} & \colhead{Mg$^b$} & \colhead{Fe5270} &
  \colhead{Fe5335} & \colhead{$\sigma_0$} & \colhead{$\left< \mu_V
  \right>$} & \colhead{$R_e$} & \colhead{$S/N$}\\
\colhead{} & \colhead{($\mbox{\AA}$)} & \colhead{($\mbox{\AA}$)} &
  \colhead{($\mbox{\AA}$)} & \colhead{($\mbox{\AA}$)} &
  \colhead{($\mbox{\AA}$)} & \colhead{($\mbox{\AA}$)} & \colhead{(km
  s$^{-1}$)} & \colhead{(mag $\arcsec^{-2}$)} & \colhead{($\arcsec$)}
  & \colhead{$\mbox{\AA}^{-1}$ Observed}}

\startdata
p0i102c3 &  -0.3 &  -0.1 &  -4.1 &  4.7 &  2.2 &  2.4 & $154\pm10$ & 21.13 &  1.3 & 23.4 \\
p0i139c3 &   0.0 &  -0.4 &  -2.8 &  4.0 &  2.8 &  2.8 & $140.\pm7$ & 18.85 &  0.3 & 16.7 \\
p0i15c3 &   0.0 &  -3.2 &  -6.3 &  4.7 &  2.2 &  1.3 & $278\pm26$ & 20.30 &  0.7 & 12.3 \\
p0i170c3 &   0.0 &  -1.3 &  -3.2 &  4.8 &  1.5 &  1.9 & $185\pm13$ &  --   &  -- & 26.6 \\
p0i1c3 &  -0.3 &   3.3 &  -1.4 &  4.0 &  2.0 &  3.3 & $105\pm19$ & 20.61 &  0.5 &  6.9 \\
p0i1c4 &  -0.5 &   0.5 &  -2.0 &  --  &  2.6 &  2.5 & $ 69\pm10$ & 19.72 &  0.4 & 11.9 \\
p0i206c3 &  -0.1 &  0.0 &  -5.6 &  1.2 &  0.7 &  1.2 & $491\pm22$ & 19.87 &  1.2 & 19.9 \\
p0i209c3 &   0.0 &  -5.0 &  -6.5 &  2.7 &  2.8 &  0.9 & -- &  --   &  --  &  1.6 \\
p0i217c3 &   0.0 &  -1.3 &  -5.6 &  4.2 &  2.3 &  1.9 & $144\pm11$ & 19.55 &  0.4 & 14.0 \\
p0i24c2 &   0.0 &   0.0 &  -6.3 &  2.2 &  3.2 &  2.7 & $165\pm24$ & 18.97 &  0.2 &  8.2 \\
p0i38c4 &   0.0 &  -0.6 &  -9.0 &  3.9 &  3.6 &  0.1 & -- &  --   &  --  &  6.5 \\
p0i39c3 &   0.0 &   0.3 &  -4.7 &  6.7 &  2.4 &  2.6 & $204\pm22$ & 20.20 &  0.8 & 16.3 \\
p0i3c3 &  0.0 &  -2.5 &  -5.6 &  4.2 &  1.8 &  1.8 & $300.\pm31$ & 19.22 &  0.5 & 24.4 \\
p0i42c3 &   0.0 &  -0.4 &  -4.7 &  4.3 &  2.9 &  2.3 & $252\pm18$ & 19.55 &  0.9 & 40.9 \\
p0i45c2 & -10.1 &   6.4 &  -5.6 &  0.1 &  3.4 &  3.3 & -- &  --   &  --  &  6.6 \\
p0i49c2 &   0.0 &  -1.6 &  -3.8 &  2.8 &  1.4 &  2.3 & $236\pm16$ & 20.75 &  1.2 & 22.5 \\
p0i4c4 &  -0.8 &  -0.6 &  -1.1 &  3.8 &  3.0 &  2.3 & $128\pm13$ & 20.28 &  0.4 &  9.5 \\
p0i53c4 &   0.0 &  -1.8 &  -2.0 &  3.2 &  1.1 &  0.9 & $403\pm32$ & 20.00 &  1.2 & 17.2 \\
p0i55c2 &   0.0 &   5.9 &   1.1 &  --  &  2.9 &  0.8 & -- &  --   &  --  &  5.8 \\
p0i66c3 &  -6.0 &  -0.3 &  -4.0 &  3.2 &  3.6 &  2.4 & $ 86\pm 7$ & 21.15 &  1.1 & 15.3 \\
p0i72c3 &   0.0 &  -1.9 &  -5.8 &  4.6 &  2.4 &  2.3 & $275\pm23$ &  --   &  --  & 42.2 \\
p0i79c3 &   0.0 &  -1.2 &  -5.4 &  5.4 &  2.1 &  2.5 & $283\pm23$ & 21.08 &  3.1 & 28.5 \\
p0i85c3 &  -1.7 &  -0.4 &  -5.4 &  2.4 &  0.5 &  0.1 & $425\pm49$ & 20.61 &  1.4 & 20.8 \\
p0i91c3 &   0.0 &  -2.2 &  -3.1 &  5.6 &  2.9 &  2.5 & $169\pm15$ & 19.01 &  0.4 & 24.8 \\
p0i95c3 &   0.0 &  -1.3 &  -6.3 &  4.9 &  2.7 &  1.9 & $221\pm24$ & 19.36 &  0.8 & 36.9 \\
p10i1c2 &  -0.4 &  -0.2 &  -4.6 &  2.3 &  2.5 &  2.8 & $214\pm13$ & 19.52 &  0.8 & 38.4 \\
p11i147c4 &   0.0 &  -1.0 &  -2.3 &  7.6 &  3.3 &  3.1 & $126\pm 7$ & 19.42 &  0.4 & 18.9 \\
p12i160c2 &  -0.7 &   6.0 &  -1.1 &  6.4 &  2.9 &  3.5 & -- &  --   &  --  &  6.0 \\
p12i160c3 &   0.0 &   0.4 &   0.6 &  6.5 &  3.4 &  3.0 & $138\pm19$ & 19.84 &  0.6 &  8.7 \\
p12i168c4 & -28.7 &   1.4 &   2.2 &  2.3 &  2.3 &  4.0 & -- &  --   &  --  &  4.8 \\
p12i73c4 &  -2.3 &  -0.2 &  -3.2 &  --  &  --  &  --  & $132\pm12$ & 20.07 &  0.6 & 12.3 \\
p13i130c3 & -38.9 &   1.1 & -- &  --  & -0.3 &  7.2 & -- &  --   &  --  &  2.1 \\
p13i133c4 &   0.0 &  -0.3 &  -4.5 &  --  &  --  &  --  & -- &  --   &  --  &  4.0 \\
p13i135c4 &  -2.7 &  -0.6 &  -3.6 &  3.4 &  1.8 &  2.2 & $143\pm11$ & 21.21 &  1.6 & 17.1 \\
p13i1c4 &  -1.4 &  -3.0 &  -5.7 &  8.4 &  --  &  --  & $143\pm22$ & 20.37 &  0.8 &  8.4 \\
p13i25c2 &   0.0 &  -0.3 &  -5.0 &  1.1 &  2.6 &  2.8 & $147\pm 8$ & 19.74 &  0.6 & 23.4 \\
p13i78c4 &  -0.2 &  -0.1 &  -1.6 &  --  &  2.5 &  2.0 & $232\pm21$ & 19.72 &  0.8 & 19.6 \\
p13i86c4 & -25.1 &   7.5 &  -0.7 &  --  &  --  &  --  & -- &  --   &  --  &  2.5 \\
p14i2c2 &  -7.7 &   2.0 &  -0.4 &  6.1 &  1.4 &  1.5 & $119\pm13$ & 21.22 &  0.8 & 12.5 \\
p15i144c4 &  -0.8 &   1.9 &  -5.0 &  --  &  0.7 &  2.9 & $130.\pm11$ & 19.71 &  0.5 & 12.4 \\
p17i2c4 &  -3.9 &  -0.9 &   0.6 &  2.7 &  3.4 &  5.2 & -- &  --   &  --  &  6.4 \\
p18i43c2 &   0.0 &  -0.9 &  -4.4 &  4.3 &  1.8 &  2.4 & $234\pm21$ & 19.94 &  1.0 & 32.6 \\
p18i51c2 &  0.0 &  -0.1 &   0.4 &  --  &  2.1 &  1.5 & $152\pm10$ & 18.93 &  0.3 & 18.9 \\
p18i66c3 &   0.0 &  -1.6 &  -4.9 &  4.2 &  2.3 &  2.0 & $168\pm14$ & 18.62 &  0.3 & 20.8 \\
p19i1c3 &  -0.7 &   0.0 &  -6.0 &  --  &  --  &  --  & -- &  --   &  --  &  7.0 \\
p19i75c4 &  -1.8 &   2.7 &  -0.5 &  5.4 &  0.0 &  1.6 & -- &  --   &  --  &  5.0 \\
p19i92c2 & -49.9 &   3.4 &  -1.8 &  2.2 & -0.6 &  0.4 & -- &  --   &  --  &  6.8 \\
p20i35c4 &  -6.5 &   0.3 &  -0.7 &  1.6 &  1.8 &  2.0 & $175\pm11$ & 19.41 &  0.5 & 25.1 \\
p20i48c3 &   0.0 &  -1.3 &  -4.0 &  3.9 &  2.5 &  0.5 & $172\pm11$ & 19.23 &  0.4 & 10.0 \\
p24i115c3 & -43.1 &   3.8 & -- &  9.0 &  1.2 & -0.2 & -- &  --   &  --  &  4.5 \\
p24i1c3 &  -0.5 &   2.8 &   1.1 &  1.9 & -0.7 &  0.9 & $144\pm16$ & 20.44 &  1.2 & 12.4 \\
p24i42c4 &   0.0 &  -4.9 &  -1.1 &  1.7 &  2.6 & -1.2 & -- &  --   &  --  &  5.6 \\
p24i79c4 &   0.0 &  -0.8 &   3.9 &  2.6 &  1.0 &  1.1 & -- &  --   &  --  &  4.8 \\
p24i87c3 &   0.0 &  -2.4 &  -6.3 &  3.8 &  1.8 &  0.7 & $153\pm 8$ & 19.75 &  0.7 & 11.0 \\
p25i29c3 &  -0.7 &  -0.2 &  -2.5 &  3.2 &  2.8 &  2.9 & $152\pm16$ & 18.36 &  0.2 & 13.4 \\
p25i93c3 & -30.1 &  -1.2 &  -2.3 &  2.8 &  3.0 &  2.2 & -- &  --   &  --  &  5.5 \\
p27i174c3 &   0.0 &   0.2 &  -3.4 &  5.0 &  2.7 &  1.9 & $187\pm17$ & 19.64 &  0.8 & 24.8 \\
p27i9c2 & -29.8 &  -3.2 &   3.8 &  2.7 &  --  & -1.7 & -- &  --   &  --  &  3.9 \\
p28i87c2 & -46.5 &   2.4 &  -2.2 &  1.8 &  0.5 &  1.2 & -- &  --   &  --  &  7.4 \\
p35i112c2 &   0.0 &  -1.4 &  -0.3 &  3.9 &  2.5 &  1.7 & $ 71\pm 8$ &  --   &  --  &  9.6 \\
p35i143c2 &   0.0 &  -4.0 &  -9.6 &  5.4 &  2.8 &  2.7 & $148\pm 9$ & 18.99 &  0.4 & 19.8 \\
p35i43c3 &  -4.4 &  -0.9 &  -4.4 &  3.6 &  3.0 &  2.5 & $170.\pm12$ & 20.36 &  1.1 & 18.2 \\
p35i6c3 &   0.0 &  -1.1 &  -5.0 &  4.2 &  2.5 &  2.9 & $220\pm12$ & 19.11 &  0.7 & 27.0 \\
p35i78c2 & -17.0 &  -1.6 &  -4.5 &  3.5 &  1.4 &  2.2 & $191\pm13$ & 19.36 &  0.6 & 36.1 \\
p35i82c3 &   0.0 &  -0.7 &  -5.1 &  4.9 &  2.3 &  2.4 & $201\pm16$ & 19.62 &  0.7 & 36.4 \\
p36i106c3 &   0.0 &   3.1 &  -3.0 &  --  &  --  &  --  & -- &  --   &  --  &  3.0 \\
p36i146c3 &   0.0 &  -1.2 &   7.1 &  3.4 &  2.2 &  3.0 & $167\pm12$ & 20.15 &  0.7 & 12.5 \\
p36i182c2 &   0.0 &  -0.6 &  -4.4 &  2.6 &  2.2 &  2.4 & $130.\pm 6$ & 20.09 &  0.9 & 26.6 \\
p36i182c3 &  -6.6 &   2.4 &  -1.0 &  --  &  7.6 &  1.5 & $108\pm24$ & 20.93 &  0.7 &  6.8 \\
p36i184c2 &  -0.5 &  -2.5 &  -3.4 &  4.9 &  1.6 &  2.7 & $165\pm10$ & 20.00 &  0.9 & 27.2 \\
p36i187c3 &  -2.9 &  -0.2 &   4.9 &  4.4 &  1.3 &  3.1 & $123\pm10$ & 19.80 &  0.5 &  8.8 \\
p36i1c4 &  -3.3 &  -0.6 &  -4.2 &  4.1 &  2.9 &  2.5 & $ 79\pm10$ & 19.12 &  0.5 & 22.2 \\
p36i2c2 &   0.0 &  -1.7 &  -5.1 &  2.0 &  2.1 &  1.6 & $376\pm24$ & 20.42 &  1.8 & 22.6 \\
p36i2c3 &  -2.6 &   1.3 &   6.5 &  6.8 &  3.0 &  2.5 & $164\pm10$ & 21.76 &  1.5 & 16.2 \\
p36i31c4 &  0.0 &  -3.1 &  -4.1 &  4.4 &  2.0 &  1.6 & $264\pm28$ & 20.14 &  1.1 & 32.6 \\
p36i32c4 &   0.0 &  -3.1 &  -3.4 &  4.5 &  2.9 &  3.1 & $215\pm19$ & 19.48 &  0.4 &  7.6 \\
p36i34c2 &   0.0 &  -1.1 &  -5.0 &  --  &  1.7 &  2.6 & $114\pm10$ & 19.52 &  0.3 & 10.4 \\
p36i36c4 &  -5.9 &   0.0 &  -9.5 &  4.7 &  1.3 &  0.8 & -- &  --   &  --  &  5.8 \\
p36i3c2 &   0.0 &  -0.4 &  -8.9 &  2.3 &  2.6 &  2.2 & $110.\pm 9$ & 19.64 &  0.5 & 15.3 \\
p36i48c3 &  -1.4 &  -4.1 &  -8.1 &  7.1 &  1.9 &  2.1 & $188\pm15$ & 19.97 &  0.6 & 12.8 \\
p36i5c2 &  -6.8 &   2.1 &  -2.1 &  0.3 &  2.5 &  1.7 & $169\pm15$ & 20.41 &  1.2 & 12.5 \\
p36i67c4 & -14.3 &   1.0 &  -0.8 &  4.4 &  --  &  --  & -- &  --   &  --  &  7.0 \\
p36i70c4 &  -8.7 &   2.0 &  -0.9 &  2.3 &  2.6 &  1.6 & $ 83\pm 8$ & 19.55 &  0.4 & 18.5 \\
p36i71c3 &  -4.7 &   1.1 &  -4.2 &  4.3 &  1.3 &  1.4 & -- &  --   &  --  &  7.3 \\
p36i76c2 &  -4.5 &   0.1 &   0.7 &  4.3 &  2.8 &  2.7 & $261\pm13$ & 20.12 &  0.5 & 13.2 \\
p36i76c3 &  -2.8 &   3.0 &  -2.0 &  5.7 &  0.3 &  0.9 & $ 87\pm 14$ & 20.93 &  1.1 &  9.8 \\
p36i89c2 &  -4.0 &   1.2 &   0.4 &  --  &  2.1 &  2.2 & $107\pm10$ & 21.14 &  1.1 & 14.9 \\
p37i110c2 &   0.0 &   2.7 &  -3.5 &  7.3 &  3.2 &  3.2 & $114\pm 8$ & 19.67 &  0.3 & 10.5 \\
p37i122c2 & -41.3 &   0.8 &  -3.4 &  2.0 &  1.7 &  1.4 & -- &  --   &  --  &  4.0 \\
p37i16c4 &  -1.6 &   1.5 &  -2.5 &  3.7 &  2.2 &  1.9 & $233\pm11$ & 19.86 &  1.4 & 55.1 \\
p37i206c4 &   0.0 &   0.5 &  -7.3 &  --  &  2.7 &  2.7 & $ 64\pm 6$ & 19.12 &  0.3 & 14.5 \\
p37i207c4 &   0.0 &  -0.8 &  12.5 &  7.8 &  2.2 &  1.8 & -- &  --   &  --  &  1.8 \\
p37i64c4 &  -1.3 &   1.1 &   1.2 &  4.6 &  1.7 & -2.3 & $254\pm45$ & 18.86 &  0.3 &  7.1 \\
p37i9c4 &  -5.9 &   2.1 &  -1.3 &  0.4 &  3.1 &  1.2 & -- &  --   &  --  & 10.1 \\
p38i1c4 &  -0.9 &  -2.9 &  -0.4 &  --  & -0.8 &  2.1 & -- &  --   &  --  &  5.5 \\
p38i76c4 &   0.0 &   5.6 &  -8.5 &  --  &  0.5 &  1.6 & -- &  --   &  --  &  6.7 \\
p3i72c2 &  -7.9 &   0.5 &  -6.8 &  0.8 &  6.9 &  5.9 & -- &  --   &  --  &  2.8 \\
p5i9c3 &  -0.7 &  -0.3 &  -1.9 &  4.2 &  2.0 &  2.3 & $ 100.\pm10.$ & 18.93 &  0.2 & 14.3 \\
p7i128c3 &   0.0 &  -0.8 &  -5.0 &  3.4 &  2.3 &  2.5 & $204\pm14$ &  --   &  --  & 27.2 \\
p7i133c3 &   0.0 &  -1.3 &  -3.6 &  --  &  3.4 & -0.6 & $161\pm21$ & 21.12 &  0.7 &  7.7 \\
p7i136c3 &   0.0 &   0.2 &  -4.4 &  2.3 &  3.4 &  1.9 & $168\pm10$ & 20.08 &  0.7 & 18.4 \\
p7i1c2 &  -6.4 &  -1.5 &  -6.9 &  3.6 &  2.2 &  2.5 & $222\pm14$ & 19.75 &  1.0 & 46.8 \\
p7i67c3 &   0.0 &   1.0 &  -3.6 &  9.9 &  2.6 &  3.2 & $127\pm 8$ & 19.44 &  0.3 & 16.5 \\
p9i20c3 &  -6.1 &   0.3 &  -3.6 &  3.3 &  3.2 &  2.1 & $112\pm10$ & 21.29 &  1.6 & 22.3 \\
\enddata
\tablecomments{~Line strengths are given in $\mbox{\AA}$ of equivalent
  width, with negative values denoting emission, and positive values
  indicating absorption. No aperture corrections have been applied to
  the indices. $\sigma_0$ lists velocity dispersions, when
  measured, and the values are aperture-corrected to a $3\farcs4$
  diameter aperture at the distance to the Coma cluster. $<\mu_V>$
  indicates the mean surface brightness within the effective radius,
  $R_e$, in rest frame V-band, and is corrected for cosmological
  dimming. Typical
  errors on line indices are less than $\pm 0.2 \mbox{\AA}$ on H$\delta_A$
  and H$\gamma_A$, and $\pm 0.1 \mbox{\AA}$ on [OII], $Mg^b$, Fe5270,
  and Fe5335. Errors on $\mu_V$ and R$_e$ are 0.1 mag and $0\farcs1$, respectively.}




\end{deluxetable*}

\end{document}